\documentclass[twocolumn]{aastex61}
\usepackage{epsfig}
\usepackage{epstopdf}
\usepackage{booktabs}
\usepackage{amsmath}

\shorttitle{TESS ACWG Follow-up Strategy}
\shortauthors{}

\begin{document}

\title{A Framework for Prioritizing the \textit{TESS} Planetary Candidates Most Amenable to Atmospheric Characterization}

\correspondingauthor{Eliza M.-R.\ Kempton}
\email{ekempton@astro.umd.edu}

\author{Eliza M.-R.\ Kempton}
\affil{Department of Astronomy, University of Maryland, College Park, MD 20742, USA}
\affil{Department of Physics, Grinnell College, 1116 8th Avenue, Grinnell, IA 50112, USA}

\author{Jacob L.\ Bean}
\affil{Department of Astronomy \& Astrophysics, University of Chicago, 5640 S.\ Ellis Avenue, Chicago, IL 60637, USA}

\author{Dana R.\ Louie}
\affil{Department of Astronomy, University of Maryland, College Park, MD 20742, USA}

\author{Drake Deming}
\affil{Department of Astronomy, University of Maryland, College Park, MD 20742, USA}

\author{Daniel D.\ B.\ Koll}
\affil{Department of Earth, Atmospheric, and Planetary Sciences, Massachusetts Institute of Technology, Cambridge, MA 02139, USA}

\author{Megan Mansfield}
\affil{Department of Geophysical Sciences, University of Chicago, 5734 S.\ Ellis Avenue, Chicago, IL 60637, USA}

\author{Jessie L.\ Christiansen}
\affil{Caltech/IPAC-NASA Exoplanet Science Institute, MC 100-22 Pasadena CA 91125}

\author{Mercedes L\'opez-Morales}
\affil{Harvard-Smithsonian Center for Astrophysics, 60 Garden Street, Cambridge, MA 01238, USA}

\author{Mark R.\ Swain}
\affil{Jet Propulsion Laboratory, California Institute of Technology, 4800 Oak Grove Drive, Pasadena, CA 91109, USA}

\author{Robert T.\ Zellem}
\affil{Jet Propulsion Laboratory, California Institute of Technology, 4800 Oak Grove Drive, Pasadena, CA 91109, USA}

\author{Sarah Ballard}
\affil{Kavli Institute for Astrophysics and Space Research, Massachusetts Institute of Technology, 77 Massachusetts Ave., Cambridge, MA 02139, USA}
\affil{MIT Torres Fellow for Exoplanetary Science}

\author{Thomas Barclay}
\affil{NASA Goddard Space Flight Center, 8800 Greenbelt Road, Greenbelt, MD 20771, USA}
\affil{University of Maryland, Baltimore County, 1000 Hilltop Cir, Baltimore, MD 21250, USA}

\author{Joanna K. Barstow}
\affil{Department of Physics and Astronomy, University College London, London, UK} 

\author{Natasha E.\ Batalha}
\affil{Space Telescope Science Institute, 3700 San Martin Dr., Baltimore, MD 21218, USA}

\author{Thomas G.\ Beatty}
\affiliation{Department of Astronomy \& Astrophysics, The Pennsylvania State University, 525 Davey Lab, University Park, PA 16802, USA}
\affiliation{Center for Exoplanets and Habitable Worlds, The Pennsylvania State University, 525 Davey Lab, University Park, PA 16802, USA}

\author{Zach Berta-Thompson} 
\affil{Department of Astrophysical and Planetary Sciences, University of Colorado, Boulder, CO 80309, USA}

\author{Jayne Birkby}
\affil{Anton Pannekoek Institute for Astronomy, University of Amsterdam, Science Park 904, 1098 XH Amsterdam, The Netherlands}

\author{Lars A.\ Buchhave}
\affil{DTU Space, National Space Institute, Technical University of Denmark, Elektrovej 328, DK-2800 Kgs. Lyngby, Denmark}

\author{David Charbonneau}
\affil{Harvard-Smithsonian Center for Astrophysics, 60 Garden Street, Cambridge, MA 01238, USA}

\author{Nicolas B.\ Cowan}
\affiliation{Department of Physics, McGill University, 3600 rue University, Montr\'eal, QC, H3A 2T8, CAN}
\affiliation{Department of Earth \& Planetary Sciences, McGill University, 3450 rue University, Montreal, QC, H3A 0E8, CAN}

\author{Ian Crossfield}
\affil{Kavli Institute for Astrophysics and Space Research, Massachusetts Institute of Technology, 77 Massachusetts Ave., Cambridge, MA 02139, USA}

\author{Miguel de Val-Borro}
\affil{NASA Goddard Space Flight Center, Astrochemistry Laboratory, 8800 Greenbelt Road, Greenbelt, MD 20771, USA}
\affil{Department of Physics, Catholic University of America, Washington, DC 20064, USA}

\author{Ren\'e Doyon}
\affil{Institut de Recherche sur les Exoplan\`etes, D\'epartment de Physique, Universit\'e de Montr\'eal, Montr\'eal, QC H3C 3J7, Canada}

\author{Diana Dragomir}
\affil{Kavli Institute for Astrophysics and Space Research, Massachusetts Institute of Technology, 77 Massachusetts Ave., Cambridge, MA 02139, USA}
\affil{NASA Hubble Fellow}

 \author{Eric Gaidos}
\affil{Department of Geology \& Geophysics, University of Hawai`i at M\"{a}noa, Honolulu, HI 96822}

\author{Kevin Heng}
\affil{University of Bern, Center for Space and Habitability, Gesellschaftsstrasse 6, CH-3012, Bern, Switzerland}

\author{Renyu Hu}
\affil{Jet Propulsion Laboratory, California Institute of Technology, 4800 Oak Grove Drive, Pasadena, CA 91109, USA}

\author{Stephen R.\ Kane}
\affil{Department of Earth Sciences, University of California, Riverside, CA 92521, USA}

\author{Laura Kreidberg}
\affil{Harvard Society of Fellows 78 Mt. Auburn St.\ Cambridge, MA 02138, USA}
\affil{Harvard-Smithsonian Center for Astrophysics, 60 Garden Street, Cambridge, MA 01238, USA}

\author{Matthias Mallonn}
\affil{Leibniz-Institut f\"{u}r Astrophysik Potsdam, An der Sternwarte 16, D-14482 Potsdam, Germany}

\author{Caroline V. Morley}
\affil{Department of Astronomy, Harvard University, 60 Garden St, Cambridge MA 02138} 

\author{Norio Narita}
\affil{Department of Astronomy, The University of Tokyo, 7-3-1 Hongo, Bunkyo-ku, Tokyo 113-0033, Japan}
\affil{JST, PRESTO, 7-3-1 Hongo, Bunkyo-ku, Tokyo 113-0033, Japan}
\affil{Astrobiology Center, NINS, 2-21-1 Osawa, Mitaka, Tokyo 181-8588, Japan}
\affil{National Astronomical Observatory of Japan, NINS, 2-21-1 Osawa, Mitaka, Tokyo 181-8588, Japan}
\affil{Instituto de Astrof¥'{i}sica de Canarias (IAC), 38205 La Laguna, Tenerife, Spain}

\author{Valerio Nascimbeni}
\affil{Dipartimento di Fisica e Astronomia, ``G. Galilei'', Universit\`a degli Studi di Padova, Vicolo dell'Osservatorio 3, I-35122 Padova, Italy}

\author{Enric Pall\'e}
\affil{Instituto de Astrof\'\i sica de Canarias (IAC), 38205 La Laguna, Tenerife, Spain}
\affil{Departamento de Astrof\'\i sica, Universidad de La Laguna (ULL), 38206, La Laguna, Tenerife, Spain}

\author{Elisa V.\ Quintana}
\affil{NASA Goddard Space Flight Center, 8800 Greenbelt Road, Greenbelt, MD 20771, USA}

\author{Emily Rauscher}
\affil{Department of Astronomy, University of Michigan, 1085 S. University Ave., Ann Arbor, MI 48109, USA}

\author{Sara Seager}
\affil{Massachusetts Institute of Technology, Department of Earth, Atmospheric, and Planetary Sciences, 77 Massachusetts Avenue, Cambridge, MA 02139, USA}
\affil{Massachusetts Institute of Technology, Department of Physics, 77 Massachusetts Avenue, Cambridge, MA 02139, USA}

\author{Evgenya L.\ Shkolnik}
\affil{School of Earth and Space Exploration, Arizona State University, Tempe, AZ, 85287}

\author{David K.\ Sing}
\affil{Astrophysics Group, Physics Building, Stocker Road, University of Exeter, Devon EX4 4QL, UK}

\author{Alessandro Sozzetti}
\affil{INAF - Osservatorio Astrofisico di Torino, Via Osservatorio 20, 10025 Pino Torinese, Italy}

\author{Keivan G.\ Stassun} 
\affil{Vanderbilt University, Department of Physics \& Astronomy, 6301 Stevenson Center Ln., Nashville, TN  37235, USA}

\author{Jeff A.\ Valenti}
\affil{Space Telescope Science Institute, 3700 San Martin Dr., Baltimore, MD 21218, USA}  

\author{Carolina von Essen}
\affil{Stellar Astrophysics Centre, Department of Physics and Astronomy, Aarhus University, Ny Munkegade 120, DK-8000 Aarhus C, Denmark}

\begin{abstract}

A key legacy of the recently launched \textit{TESS} mission will be to provide the astronomical community with many of the best transiting exoplanet targets for atmospheric characterization.  However, time is of the essence to take full advantage of this opportunity.  \textit{JWST}, although delayed, will still complete its nominal five year mission on a timeline that motivates rapid identification, confirmation, and mass measurement of the top atmospheric characterization targets from \textit{TESS}.  Beyond \textit{JWST}, future dedicated missions for atmospheric studies such as \textit{ARIEL} require the discovery and confirmation of several hundred additional sub-Jovian size planets ($R_{p} < 10$~$R_{\oplus}$) orbiting bright stars, beyond those known today, to ensure a successful statistical census of exoplanet atmospheres.  Ground-based ELTs will also contribute to surveying the atmospheres of the transiting planets discovered by \textit{TESS}.  Here we present a set of two straightforward analytic metrics, quantifying the expected signal-to-noise in transmission and thermal emission spectroscopy for a given planet, that will allow the top atmospheric characterization targets to be readily identified among the \textit{TESS} planet candidates.  Targets that meet our proposed threshold values for these metrics would be encouraged for rapid follow-up and confirmation via radial velocity mass measurements.  Based on the catalog of simulated \textit{TESS} detections by Sullivan et al.~(2015), we determine appropriate cutoff values of the metrics, such that the \textit{TESS} mission will ultimately yield a sample of $\sim300$ high-quality atmospheric characterization targets across a range of planet size bins, extending down to Earth-size, potentially habitable worlds.  

\end{abstract}

\section{Introduction}

The \textit{Transiting Exoplanet Survey Satellite} (\textit{TESS}) is poised to revolutionize the exoplanet field by completing the census of close-in transiting planets orbiting the nearest stars \citep{ricker15}.  The planets discovered by \textit{TESS} will be among the best targets for atmospheric characterization, owing to the large signal-to-noise (S/N) obtainable based on the brightness and relatively small sizes of their host stars.  Most notably, \textit{TESS} is designed to detect many hundreds of sub-Jovian size planets that are substantially better atmospheric characterization targets than those detected by the \textit{Kepler} satellite \citep[e.g.,][]{thompson18}. The atmospheric characterization studies enabled by the \textit{TESS} mission will round out our understanding of exoplanet atmospheres in the Neptune- down to Earth-size regime and may even extend to habitable worlds.

The backdrop for the \textit{TESS} mission is the dramatic increase in our capability to probe the atmospheres of transiting exoplanets that is expected over the coming decade. The launch of the \textit{James Webb Space Telescope} (\textit{JWST}) is highly anticipated, and construction of the next generation of extremely large telescopes (ELTs) on the ground is already underway. A primary science driver for both \textit{JWST} and the ELTs is the characterization of exoplanet atmospheres\footnote{\url{https://jwst.nasa.gov/science.html}}\footnote{\url{http://www.gmto.org/Resources/GMT-SCI-REF-00482\_2\_GMT\_Science\_Book.pdf}}\footnote{\url{https://www.eso.org/sci/facilities/eelt/science/doc/eelt\_sciencecase.pdf}}\footnote{\url{https://www.tmt.org/page/exoplanets}}, and the planets discovered by the \textit{TESS} mission will be vital to realizing this vision.

\begin{figure}
\begin{center}
\includegraphics[scale=0.5]{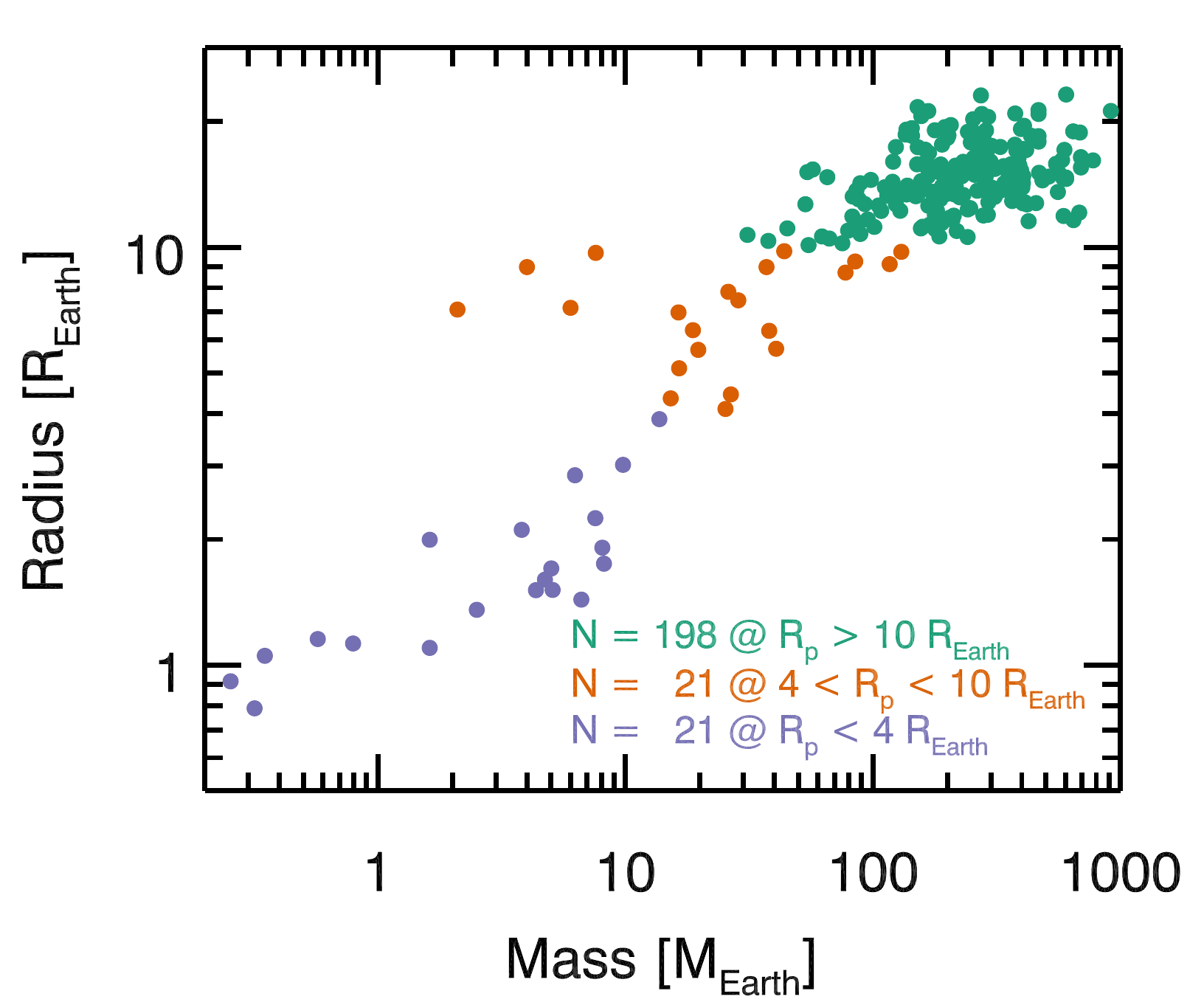}
\caption{\label{fig:known} Masses and radii of known transiting exoplanets that are good targets for atmospheric characterization. Formally, these are selected as the known planets with estimated transmission metric values greater than 50, as defined in \S\ref{analysis}.}
\end{center}
\end{figure}

Figure~\ref{fig:known} shows the known transiting exoplanets that are favorable targets for atmospheric characterization based on their expected S/N for transmission spectroscopy\footnote{Throughout this paper we use the properties of known transiting exoplanets and their host stars from TEPCat: \url{http://www.astro.keele.ac.uk/jkt/tepcat/}.}. While there are a substantial number of giants that are good targets, there are very few known planets smaller than 10\,$R_{\oplus}$ that are suitable for this type of atmospheric study. The atmospheric characterization community has the ambition to use \textit{JWST} and the ELTs to characterize many tens of planets, including a push toward temperate and rocky worlds \citep{cowan15,snellen13,rodler14}. Furthermore, hundreds of planets over a wide range of masses and radii will ultimately be needed for future dedicated exoplanet atmosphere missions like the recently selected \textit{ARIEL} concept \citep[][see \S\ref{sample} for more discussion]{tinetti16}.

While \textit{TESS} is positioned to deliver the planets needed for atmospheric characterization efforts, substantial follow-up work is needed to turn its detected planet candidates into bona fide planets that are appropriate targets for atmospheric observations. Following \textit{TESS} detection, the essential follow-up steps for this process include improved characterization of the host star \citep[e.g.,][]{stassun18}, additional transit observations to increase the precision on the orbital ephemerides \citep[e.g.,][]{kane09}, validation or confirmation of planetary nature, and planet mass measurement.

Typically the most resource-intensive component of candidate follow up is the radial velocity (RV) confirmation and measurement of planet mass.  While planet validation techniques exist that bypass this step and do not require RV mass measurements \citep{torres11}, precise mass determinations have been shown to be fundamental to correctly interpreting atmospheric observations of exoplanet transmission spectra \citep{bat17}. Furthermore, well constrained radial velocity orbits are needed to predict secondary eclipse times. Mass measurements are a key component of the \textit{TESS} Level-1 mission requirements, which specify that 50 planets with $R_{p} < 4$~$R_{\oplus}$ have their masses measured through RV follow up \citep{ricker15}.  The recent delays of the \textit{JWST} launch have afforded the exoplanet community with an additional time cushion for candidate follow-up efforts. Nevertheless, the community still must act quickly to ensure that the best atmospheric characterization targets are confirmed and weighed on the timeline of the prime \textit{JWST} mission.

Our goal here is to motivate a set of threshold criteria to identify the \textit{TESS} planet candidates that are expected to be most amenable to atmospheric characterization, and therefore merit rapid RV follow up.  We base our selection thresholds on our understanding of the expected mission planet yields from the simulated \textit{TESS} catalog of \citet{sul15}.  The \citet{sul15} catalog is one realization of the \textit{TESS} planet detection outcomes based on published occurrence rate statistics \citep{fressin13, dressing15} and a galactic stellar population model \citep{girardi05}.

The \textit{TESS} exoplanet yields have been re-examined more recently by a number of authors including \citet{bouma17}, who investigated strategies for an extension of the \textit{TESS} mission, \citet{ballard18}, who studied yields and multiplicities of planets orbiting M-dwarf hosts, \citet{barclay18}, who re-calculated the planetary yields using the actual \textit{TESS} Input Catalog (TIC) of target stars, \added{and \citet{huang18}, who also used the TIC stars but with updated parameters from the \textit{Gaia} Data Release 2 along with improved treatment of multi-planet systems and \textit{TESS} noise systematics}.  The more recent results are mostly in line with the findings of \citet{sul15}, with a couple of differences that have implications for our current work. These differences are: 1) The M-dwarf planet occurrence rates found by \citet{ballard18} and \citet{huang18} are higher than those previously reported by up to 50\%, and (2) \citet{bouma17} and \citet{barclay18} both report an overestimation of the number of Earths and super-Earths in the \citet{sul15} work that resulted from an error in the latter's calculations.  \added{Additionally, none of these works account for the paucity of planets with $R_p \approx 1.5 R_{\oplus}$ associated with the planetary radius gap identified by \citet{fulton17} and \citet{vaneylen18}, which could lead to a small over-prediction of such planets in each of the aforementioned catalogs.}   

In this work, we employ the \citet{sul15} catalog, while noting the discrepancies with more recent simulated \textit{TESS} yield calculations.  \added{As a check for the effect of these differences, we have repeated the key steps of our analysis using the \citet{barclay18} catalog, and we discuss those outcomes in Section~\ref{sec:concl}.} Using the \citet{sul15} results for our primary analysis allows us to build on the \citet{lou18} simulations of \textit{JWST}/NIRISS transit observations of the planets in that catalog. We identify cutoffs to select the top atmospheric characterization targets based on the expected S/N of the simulated \textit{TESS} planets in transmission and emission spectroscopy.  Our methodology and threshold criteria for identifying the best atmospheric characterization targets from \textit{TESS} are described below. We concentrate mainly on \textit{JWST} observability, with the expectation that the results would be qualitatively similar for calculations done specifically for ground-based telescopes or \textit{ARIEL}.

\section{Sample Selection \label{sample}}
We consider three samples of planets for our analysis of atmospheric observability, two in terms of their observability for transmission spectroscopy and one in terms of its observability for emission spectroscopy. The two transmission spectroscopy samples are: (1) a large sample of planets across a range of planet sizes and (2) a sample of small planets in and near the habitable zones of their host stars. The emission spectroscopy sample is composed of planets that have sizes consistent with a terrestrial composition. \added{The properties of} these samples are described in more detail in the following subsections, \added{and then appropriate threshold criteria to deliver these samples are identified in Section~\ref{sec:results}}.

\subsection{Statistical Sample}
Based on simulations performed for the \textit{FINESSE} mission proposal \citep{bean17}, we postulate that one goal of atmospheric characterization efforts over the next decade should be a transmission spectroscopy survey of approximately 500 planets so that statistical trends can be revealed. These simulations indicate that on order of 500 planets are needed to accurately discern population-wide properties like the relationship between atmospheric metallicity and planetary mass and differences between stellar and planetary atmospheric abundance ratios (e.g., C/O) in the face of the diversity predicted by planet formation models \citep{fortney13,mordasini16}. We focus first on transmission spectroscopy because this is expected to be the prime mode for exoplanet atmospheric observations and provides the best sensitivity to a wide range of planets. 

To create a 500 planet statistical sample we will need roughly 300 new planets from \textit{TESS} that sample the radius parameter space $R_{p}\,<\,10\,R_{\oplus}$. We therefore take the catalog of simulated \textit{TESS} detections from \citet{sul15} and divide it into the following four planet-size bins: 
\begin{itemize}
\item Terrestrials: $R_{p} < 1.5$~$R_{\oplus}$
\item Small sub-Neptunes: $ 1.5 < R_{p} < 2.75$~$R_{\oplus}$
\item Large sub-Neptunes: $ 2.75 < R_{p} < 4.0$~$R_{\oplus}$
\item Sub-Jovians: $ 4.0 < R_{p} < 10$~$R_{\oplus}$
\end{itemize}

We \added{aim to} select the best simulated planets from each of the four planet size bins to be considered for RV follow up and eventual atmospheric characterization (the latter assuming the RV mass and stellar activity metrics render the candidate a high-quality target for transmission spectroscopy). We \added{initially identify high-quality atmospheric characterization targets by using} the predicted transmission spectroscopy S/N according to the results of \citet{lou18} from their end-to-end \textit{JWST}/NIRISS simulator for the case of a 10-hour observing campaign. The \citet{lou18} S/N values are calculated for the detection of spectral features (i.e., difference from a flat line) integrated over the entire NIRISS bandpass, which is 0.6--2.8\,$\mu$m for most targets and 0.8--2.8\,$\mu$m for bright targets that require the use of a smaller subarray.

We concentrate on NIRISS because a precise simulation of the \citet{sul15} catalog has already been done for this instrument, and because NIRISS gives more transmission spectroscopy information per unit of observing time compared to the other \textit{JWST} instruments for a wide range of planets and host stars \citep{batalha17c,howe17}. However, it is worth keeping in mind that other \textit{JWST} instruments may be more suitable for observations in certain corners of parameter space. For example, \citet{morley17} suggested the use of NIRSpec for observations of small planets orbiting cool, nearby stars. Also, the predicted NIRISS S/N should be approximately scalable to the achievable S/N for ground-based observations, (see more discussion in \S\ref{stat_sample}). 

\begin{figure}
\begin{center}
\includegraphics[scale=0.5]{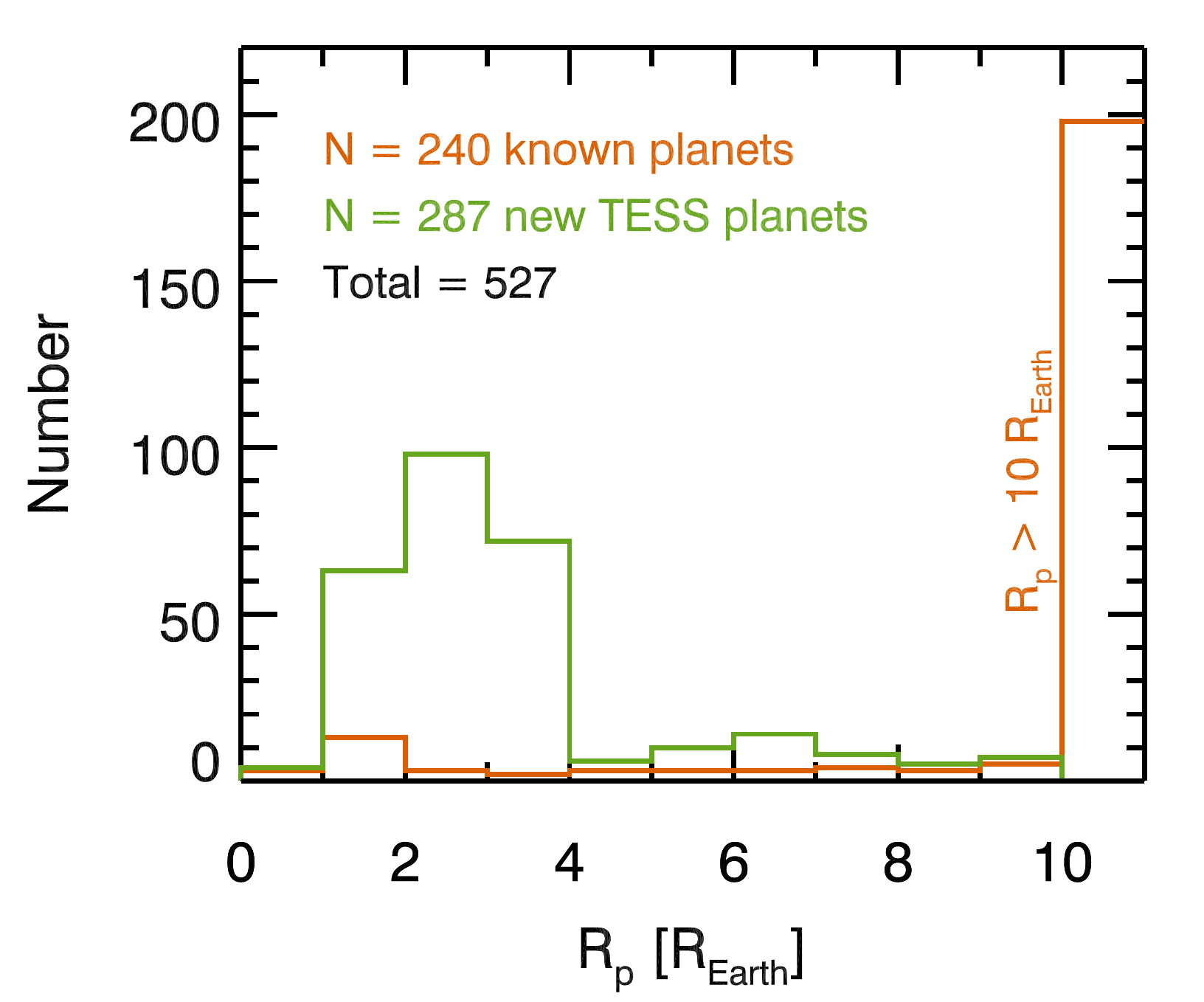}
\caption{\label{fig:histogram} Histogram of the radii for the known transiting exoplanets that are good targets for atmospheric characterization with transmission spectroscopy (orange) with the potential best \textit{TESS} planets from our statistical sample (green).}
\end{center}
\end{figure}

\added{To build a sample of $\sim 300$ total targets,} from the large and small sub-Neptune bins (total number $N = 578$ and $N = 1063$, respectively), we select the top 100 planets each. There are only 100 (exactly) planets in the sub-Jovian bin, and we ultimately recommend to follow up the best 50 (see \S\ref{stat_sample}). From the terrestrial planet bin, in which the total number of expected \textit{TESS} discoveries drops off because the transit depths approach the mission's detection threshold, we select the top quintile planets (37 out of a total $N = 192$). The combined sample of 287 planets with $R_{p}\,<\,10\,R_{\oplus}$ from the four size bins constitutes our ``statistical'' transmission spectroscopy sample. The histogram of the planet radii for the known planets (the same ones shown in Figure~\ref{fig:known}) and this statistical sample is shown in Figure~\ref{fig:histogram}.   

\subsection{Small Temperate Sample}
In addition to the statistical sample of the most easily characterizable planets of a given size, we also consider a sample of planets in and near the liquid water habitable zone as targets for transmission spectroscopy measurements. Following \citet{sul15}, we delineate this sample as being planets with insolation values of $S_{p} = 0.2 - 2.0$ times the Earth's insolation ($S_{\oplus}$) and $R_{p} < 2.0$~$R_{\oplus}$. A total of 60 simulated planets from the \citet{sul15} catalog meet the criteria of this ``small temperate'' transmission spectroscopy sample, and we perform a further down-selection of this sample based on transmission spectroscopy detectability (see \S\ref{sm_tem}).

We note that the lower insolation boundary (0.2\,$S_{\oplus}$) of the small temperate sample is commensurate with the outer edge of the habitable zone for low-mass stars as calculated by \citet{kopparapu13}, while the higher insolation boundary of this sample (2\,$S_{\oplus}$) is well interior to the inner edge ($\sim 0.9\,S_{\oplus}$, again for low-mass stars). We extend our sample to include planets substantially inward of the nominal habitable zone boundary because these planets are crucial for testing both the concept of the habitable zone \citep{bean17b} and theories of atmospheric evolution that are relevant for potentially habitable planets \citep{schaefer16,morley17}.

\subsection{Emission Sample}
\citet{morley17} have recently suggested that thermal emission measurements at long wavelengths (i.e., $\lambda\,>\,5\,\mu$m) with \textit{JWST}/MIRI could be more insightful than transmission measurements for the warmer terrestrial planets for a given observing time. Therefore, we also estimate the emission spectroscopy secondary eclipse S/N for the \citet{sul15} planets in the terrestrial bin as a special sample. There is no MIRI emission spectroscopy analogue of the \citet{lou18} paper, so we consider all the terrestrial planets in the \citet{sul15} catalog and scale the S/N calculations to the expected secondary eclipse depth for a well-studied example planet (see \S\ref{esm}). We refer to the targets in this group as the ``emission'' sample.

The choice here to focus on just terrestrial planets for emission spectroscopy is not to say that such observations are not interesting for larger planets, but rather that a full scale MIRI S/N estimate of secondary eclipses for the \citet{sul15} catalog is beyond the scope of this paper. We choose to focus on the terrestrial planets as there is a particular need to identify additional planets to target with \textit{JWST}'s unique capabilities to study small signals at long wavelengths. Furthermore, larger planets that are good emission spectroscopy targets will generally also be good targets for transmission spectroscopy because both methods have similar dependencies on planet and host star size and planet temperature, although emission spectroscopy has a much steeper dependency on the latter.

\section{Analysis  \label{analysis}}
In this section we write down analytic metrics for the expected S/N of transmission and emission spectroscopy observations. The transmission spectroscopy metric is applied to the statistical and small temperate planet samples, and the emission spectroscopy metric is applied to the terrestrial planet emission sample. 

\subsection{Transmission Metric \label{tsm}}
For each planet in the statistical and small temperate samples, we calculate a transmission spectroscopy metric (TSM) that is proportional to the expected transmission spectroscopy S/N, based on the strength of spectral features ($\propto R_{p} H / R_{*}^{2}$, where $H$ is the atmospheric scale height) and the brightness of the host star, assuming cloud-free atmospheres.  
\begin{equation}
    \textrm{TSM} = (\textrm{Scale factor}) \times \frac{R_{p}^3\ T_{eq}}{M_{p}\ R_{*}^2} \times \large{10^{-m_{J}/5}} \label{transmission_metric}
\end{equation}
The quantities in Equation~\ref{transmission_metric} are defined as follows:
\begin{itemize}
\item $R_{p}$: the radius of the planet in units of Earth radii,
\item $M_{p}$: the mass of the planet in units of Earth masses, which, if unknown, should be calculated using the empirical mass-radius relationship of \citet{chen17} as implemented by \citet{lou18},
\begin{eqnarray}
\begin{aligned}
    M_{p} & = 0.9718\,R_{p}^{3.58}\ \mathrm{for}\ R_{p} < 1.23 R_{\oplus}, \\
    M_{p} & = 1.436\,R_{p}^{1.70}\ \  \mathrm{for}\ 1.23 < R_{p} < 14.26 R_{\oplus},
\end{aligned}
\end{eqnarray}
\item $R_{*}$: the radius of the host star in units of solar radii,
\item $T_{eq}$: the planet's equilibrium temperature in Kelvin calculated for zero albedo and full day-night heat redistribution according to,
\begin{equation}
    T_{eq} =  T_{*} \sqrt{\frac{R_{*}}{a}} \left(\frac{1}{4}\right)^{1/4} \label{teq},
\end{equation}
where $T_{*}$ is the host star effective temperature in Kelvin, and $a$ is the orbital semi-major axis given in the same units as $R_{*}$,
\item $m_{J}$: the apparent magnitude of the host star in the J band, chosen as a filter that is close to the middle of the NIRISS bandpass.
\end{itemize}

The ``scale factor'' in Equation~\ref{transmission_metric} is a normalization constant selected to give one-to-one scaling between our analytic transmission metric and the more detailed work of \citet{lou18} for their 10-hour simulations \added{(with half of that time occurring during transit)}. The scale factor also absorbs the unit conversion factors so that the parameters can be in natural units. By including this normalizing factor, Equation~\ref{transmission_metric} reports near-realistic values of the expected S/N for 10-hour observing programs with \textit{JWST}/NIRISS, modulo our assumptions on atmospheric composition, cloud-free atmospheres, and a fixed mass-radius relation. We determine the scale factor separately for each planet radius bin using the average of the planets with $m_{J}\,>\,9$ (see \S\ref{stat_sample} for a discussion of the metric's applicability to bright stars). The resulting values are given in Table~\ref{table:metric}.

Our transmission metric, and its function of selecting the top atmospheric characterization targets, is similar to that of \citet{zellem17}. \citet{morgan18} have also developed a ranking metric, although there are key differences in the implementation  (e.g. the use of an explicit mass-radius relation, and no order unity correction for transit duration). 

By not including a factor of the mean molecular weight, $\mu$, in Equation~\ref{transmission_metric}, we implicitly assume that all planets in a given size bin have the same atmospheric composition.  This is the same assumption made by \citet{lou18}, who chose values of $\mu = 18$ (in units of the proton mass) for planets with $R_{p} < 1.5$~$R_{\oplus}$, and $\mu = 2.3$ for planets with $R_{p} > 1.5$~$R_{\oplus}$.  The former is consistent with a water (steam) atmosphere and the latter with a solar composition, H$_{2}$-dominated atmosphere. Any intrinsic spread in the atmospheric composition (expected primarily in the smaller planet size bins) will translate linearly into the S/N realized in an actual \textit{JWST} observing campaign.  For the calculations of the small temperate sample we re-scale the \citet{lou18} S/N and our metric for the $1.5 < R_{p} < 2.0$~$R_{\oplus}$ planets by a factor of 2.3/18 to put these planets on the same basis as the smaller planets, in terms of their mean molecular weights. That is, we assume for the purpose of investigating planetary habitability that the planets in question all have dense, secondary atmospheres.

\subsection{Emission Metric \label{esm}}
For the planets in the terrestrial bin we also compute an emission spectroscopy metric (ESM) that is proportional to the expected S/N of a \textit{JWST} secondary eclipse detection at mid-IR wavelengths. The metric \added{(which is also similar to the one from \citet{zellem18})} is

\begin{equation}
    \textrm{ESM} = 4.29\times10^6 \times \frac{B_{7.5}(T_{day})}{B_{7.5}(T_{*})} \times \left(\frac{R_p}{R_*}\right)^2 \times 10^{-m_K/5} \label{emission_metric}
\end{equation}

The new quantities in Equation~\ref{emission_metric} are defined as follows:
\begin{itemize}

\item $B_{7.5}$: Planck's function evaluated for a given temperature at a representative wavelength of 7.5\,$\mu$m,
\item $T_{day}$: the planet's dayside temperature in Kelvin, which we calculate as $1.10 \times T_{eq}$,
\item $m_{K}$: the apparent magnitude of the host star in the K band.
\end{itemize}

The second and third terms in Equation~\ref{emission_metric} provide the appropriate scaling of the secondary eclipse depth at a wavelength of 7.5~$\mu$m, which is the center of the MIRI LRS bandpass \citep[5 -- 10\,$\mu$m,][]{rieke15,kendrew15}.  The final term scales the S/N by the K-band flux of the planet's host star. We chose the K-band magnitude for the ESM because it is the longest wavelength photometric magnitude that is given in the \citet{sul15} catalog. 

The factor of $4.29\times10^6$ in front of Equation~\ref{emission_metric} scales the ESM to yield a S/N of 7.5 in the MIRI LRS bandpass for a single secondary eclipse of the reference planet GJ\,1132b \citep{berta15}, based on detailed modeling of its atmosphere described below.  We chose GJ\,1132b as the reference because \citet{morley17} have shown that it is the best known small exoplanet for thermal emission measurements with \textit{JWST}. We also confirm that many of the small \textit{TESS} planets that are likely good targets are similar to this planet. Throughout this analysis we assume the \citet{dittmann17} values for the properties of GJ\,1132b and its host star ($R_p$/$R_*$\,=\,0.0455, $a$/$R_*$\,=\,16.54, $T_*$\,=\,3270\,K). 

The emission metric assumes that both the planet and the star emit as blackbodies. For stars this tends to be a reasonable assumption at mid-IR wavelengths, although continuum H$^{-}$ opacity combined with line blanketing at shorter wavelengths results in infrared brightness temperatures that differ from the effective temperature. For example, for the benchmark planet GJ\,1132b we find that models predict that its host star's brightness temperature is 90\% of its effective temperature in the MIRI LRS bandpass (i.e., 2922 vs.\ 3270\,K). However, since this factor is different for varying stellar types we elect not to apply a correction to the ESM  beyond the normalization factor that is already applied in Equation~\ref{emission_metric}. We additionally performed tests that show that the relative scaling of planets according to the ESM in Equation~\ref{emission_metric} is not sensitive to 10\% changes in the stellar temperatures.

The assumption of blackbody emission is more problematic for the planets because their emergent mid-IR spectra are strongly sculpted by molecular absorption and the emitted flux can vary by an order of magnitude or more between spectral bands and low opacity windows that probe the deep atmosphere. However, the single temperature blackbody assumption is reasonable for a S/N metric that aims to convey the relative broadband observability of planets. We verify that our blackbody assumption gives an intermediate prediction of the emission signal for different plausible atmospheric compositions in a specific example below.

\begin{deluxetable*}{c c c c c}[htp]
\tablecaption{TSM values and associated scale factors for the statistical sample \tablenotemark{a} \label{table:metric}}
\tablehead{ 
 & \colhead{$R_{p} < 1.5$ $R_{\oplus}$} & \colhead{$1.5 < R_{p} < 2.75$ $R_{\oplus}$} & \colhead{$2.75 < R_{p} < 4.0$ $R_{\oplus}$} & \colhead{$4.0 < R_{p} < 10$ $R_{\oplus}$} }
\startdata 
First quartile (top 25) & --- & 178 & 146 & 159 \\
Second quartile (rank 25-50) & --- & 125 & 124 & \textbf{96} \\
Third quartile (rank 50-75) & --- & 109 & 95 & 51 \\
Fourth quartile (rank 75-100) & --- & \textbf{92} & \textbf{84} & 12 \\
Top quintile ($N=37$) & \textbf{12} & --- & --- & ---\\
\midrule
Scale factor & 0.190 & 1.26 & 1.28 & 1.15\\
\enddata
\tablenotetext{a}{The bold numbers indicate our suggested cutoffs for follow-up efforts.}
\end{deluxetable*}

The ESM is a stronger function of the assumed planetary temperature than the stellar temperature because observations at 7.5\,$\mu$m are nearer to the planets' peak of blackbody emission. Observations at secondary eclipse probe exoplanets' daysides. We therefore apply a theoretically-derived correction factor (1.10, see below) to the equilibrium temperature for estimating the dayside temperature needed to predict the secondary eclipse depth.

We performed theoretical calculations for the atmosphere of GJ\,1132b using 3D GCMs and 1D radiative-convective forward models to investigate the S/N scaling factor needed for the ESM (Koll et al.\ in prep). First, we used 3D GCMs to investigate the energy transport for GJ 1132b and similar synchronously rotating planets that \textit{TESS} will find. The GCM is the same as described in \citet{koll2015,koll2016}, and solves the equations of atmospheric motion coupled to gray radiative transfer. We assume the atmosphere is transparent to stellar radiation and that its infrared opacity is comparable to representative values for the Solar System \citep{robinson2014}, and we investigate the atmosphere's heat redistribution as a function of surface pressure. From the GCM calculations we find that a 1\,bar atmosphere will have moderate heat redistribution, consistent with a conventional redistribution factor of $f = 0.53$ (where $f$\,=\,1/4 is full planet redistribution and $f$\,=\,2/3 is instant re-radiation).

We also calculated 1D forward models to estimate GJ\,1132b's thermal emission signal for different compositions as a check of the ESM assumptions and to provide an absolute S/N benchmark. These include ``double-gray'' calculations of the planet's temperature-pressure profile combined with a wavelength-dependent solution of the radiative transfer equation (without scattering) to predict the planet's emission spectrum, as described in \citet{mil09}.  The calculations were done for a redistribution factor of 0.53 and a surface pressure of 1 bar for consistency with the GCM results. We also adopted an Earth-like albedo of 0.3 absent any empirical constraints on the characteristics of terrestrial exoplanet atmospheres.  This combination of albedo and redistribution results in a dayside temperature that is 10\% higher than the full redistribution equilibrium temperature calculated from Equation~\ref{teq}.  We therefore adopt a scaling of $T_{day} = 1.10 \times T_{eq}$ for our ESM calculations. 

We considered three atmospheric compositions in our 1D modeling: H$_2$-rich solar composition gas; pure H$_2$O steam; and a Venus-like composition of 96.5\% CO$_2$, 3.5\% N$_2$, plus trace amounts of H$_2$O and CO. From averaging the results for the three different types of atmospheres, we estimate a typical secondary eclipse depth of 75\,ppm for GJ\,1132b binned over the MIRI LRS bandpass. This estimate is consistent with the predictions of \citet{morley17} modulo the different assumptions of albedo and redistribution.

Finally, we estimated the noise on a single broadband secondary eclipse measurement of GJ\,1132b with MIRI LRS using the \texttt{PandExo} simulation tool \citep{batalha17}. We determined a photon-limited error of 10\,ppm from this calculation, which yields a 7.5$\sigma$ detection of GJ\,1132b according to the models described above. Our predicted significance is substantially less than that given by \citet{morley17} for similar models due to an error in those authors' calculations (Laura Kreidberg and Caroline Morley, personal communication). We also note that the \textit{JWST} throughput numbers in \texttt{PandExo} have evolved over the last year due to the incorporation of the latest instrument testing data. Our \texttt{PandExo} simulation is from February 2018; future simulations may find different results if the assumptions in \texttt{PandExo} change.

\begin{figure*}[h!]
\begin{center}
\gridline{\fig{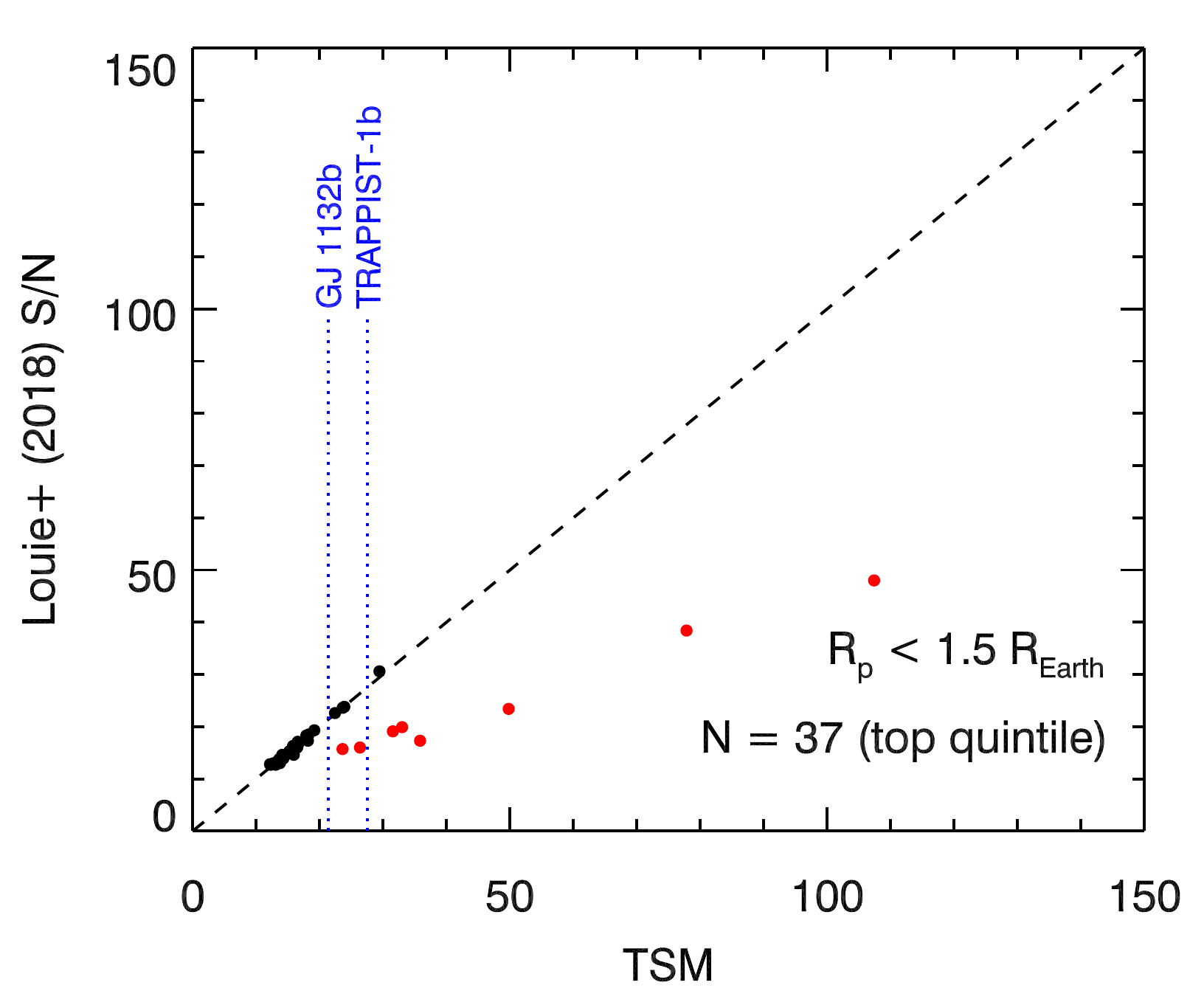}{0.49\textwidth}{}
          \fig{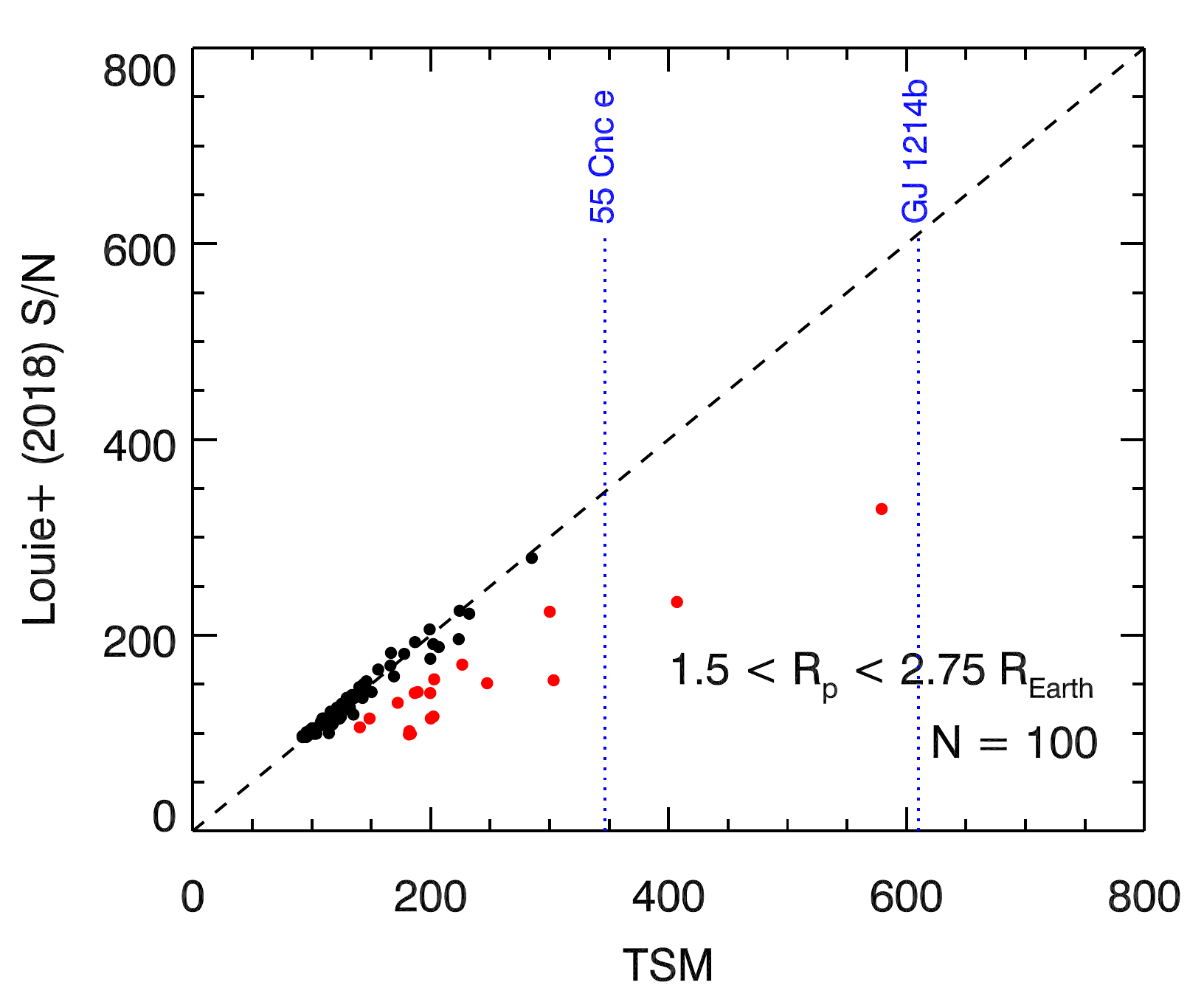}{0.49\textwidth}{}
          }
\gridline{\fig{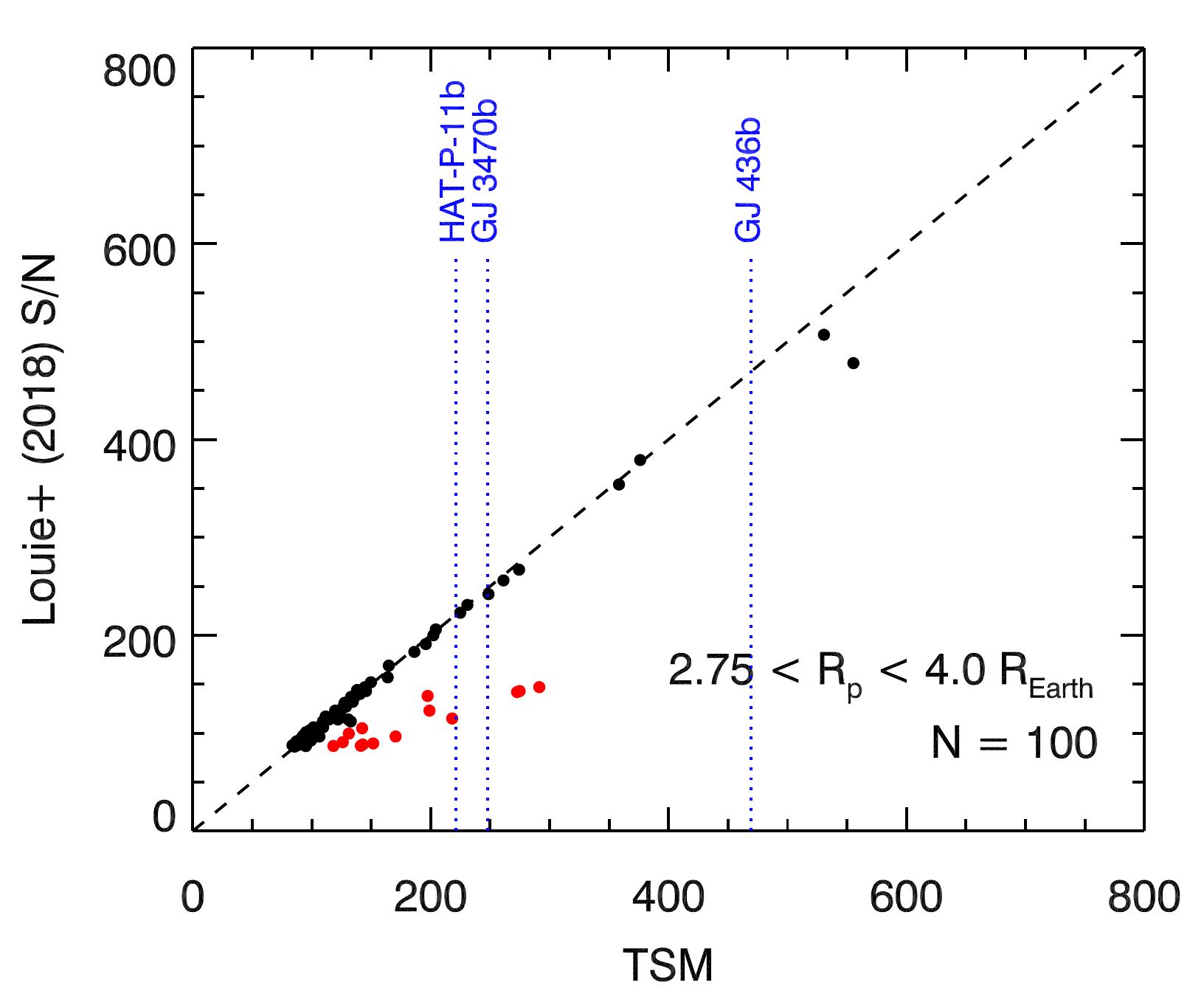}{0.49\textwidth}{}
          \fig{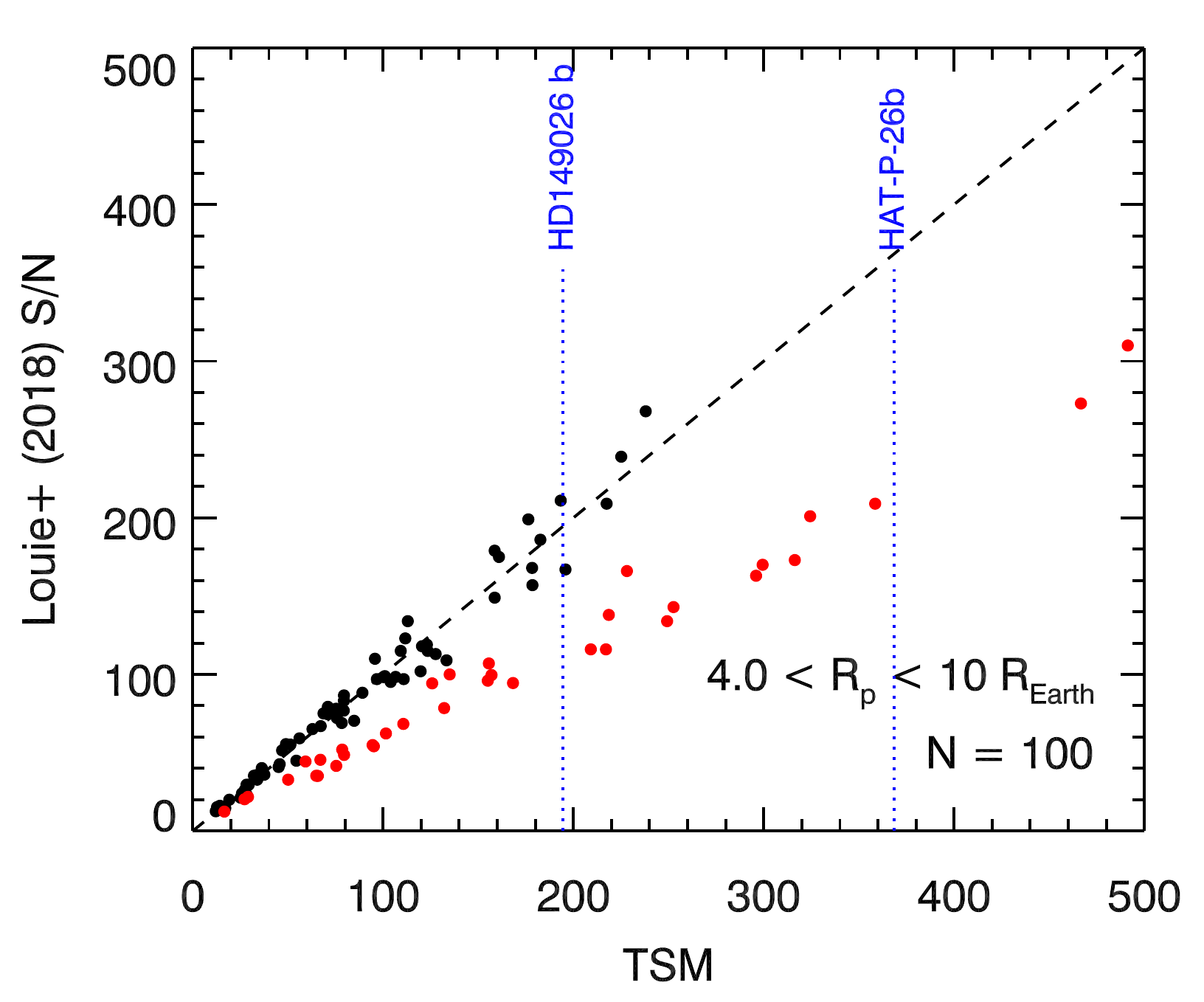}{0.49\textwidth}{}
          }
\caption{\label{fig:metric_size} NIRISS S/N from \citet{lou18} vs.\ the transmission spectroscopy metric from Equation~\ref{transmission_metric}, using the scale factors from Table~\ref{table:metric}. The points are simulated \textit{TESS} planets (black: $m_J\,>\,9$, red: $m_J\,<\,9$). The dashed line plots a one-to-one relationship. The brighter stars likely deviate from the one-to-one relationship because they have lower duty cycle for \textit{JWST} observations, which our analytic metric doesn't capture. The TSM values for known benchmark planets are indicated by the x-axis position of \added{the vertical blue lines}, assuming these planets have the same atmospheric composition as the rest of the sample. }
\end{center}
\end{figure*}

\begin{figure*}
\begin{center}
\gridline{\fig{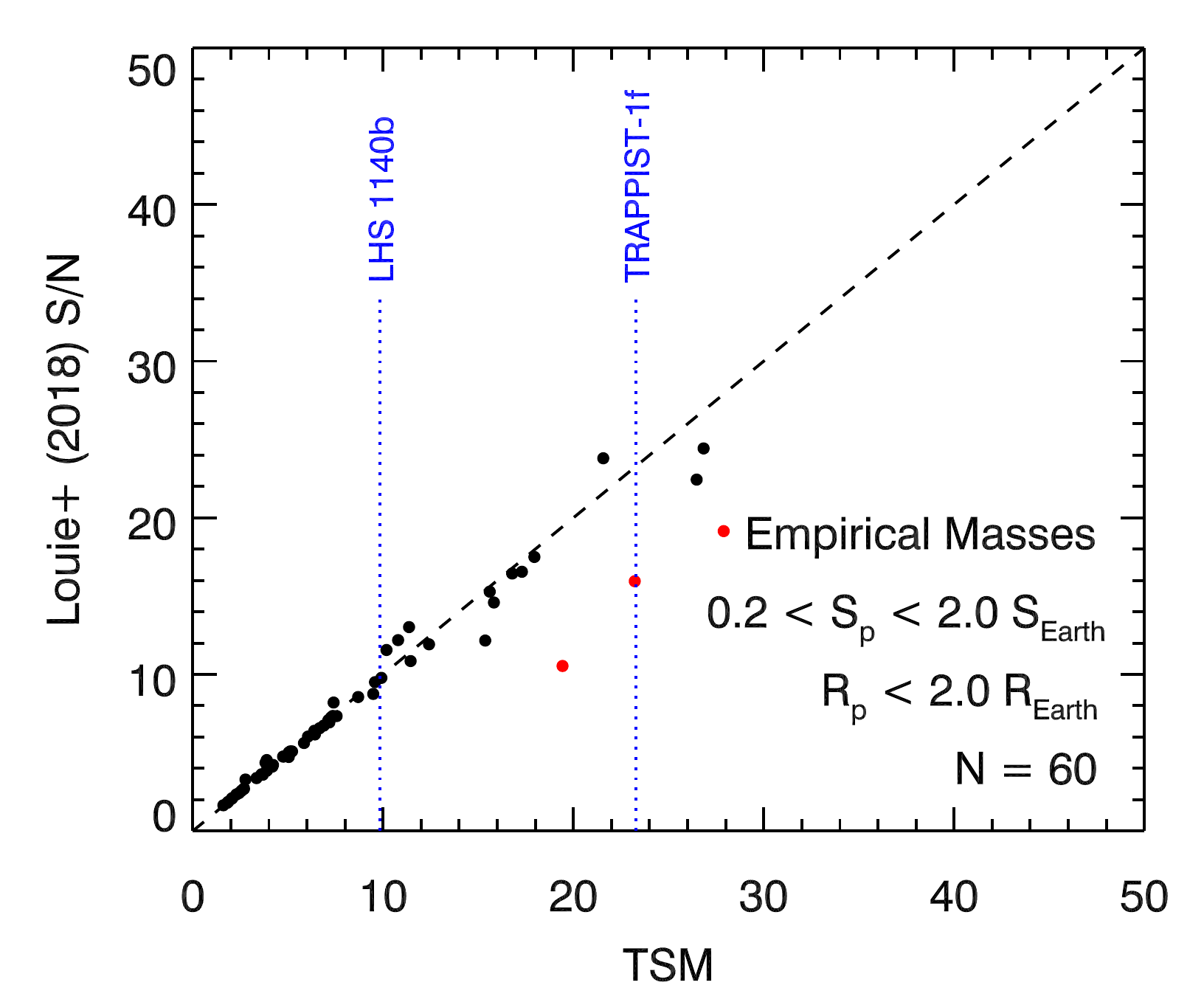}{0.49\textwidth}{}
          \fig{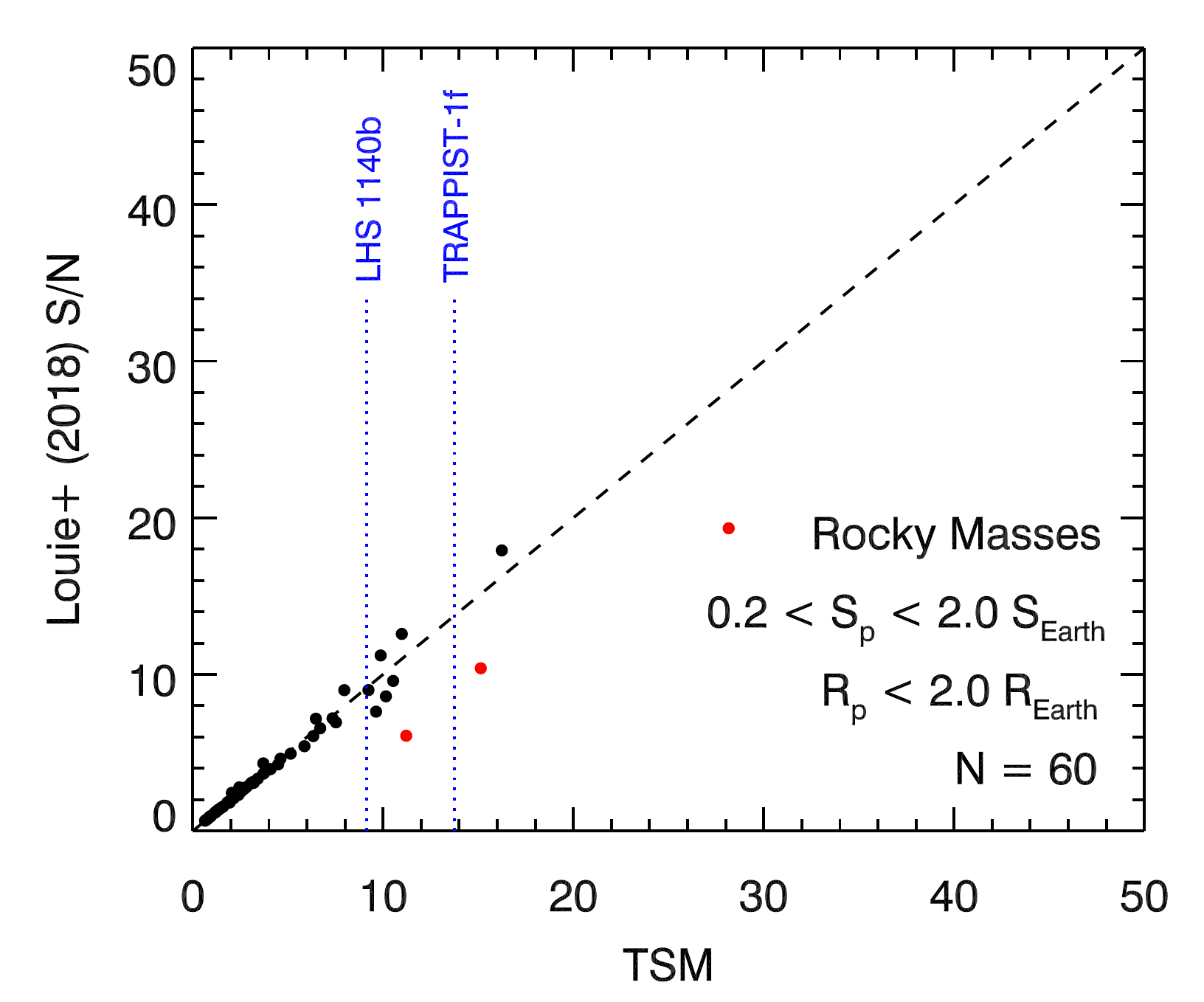}{0.49\textwidth}{}
          }
\caption{\label{fig:metric_hz} NIRISS S/N from \citet{lou18} vs.~the transmission spectroscopy metric from Equation~\ref{transmission_metric} for the small temperates, following the same labeling conventions as Figure~\ref{fig:metric_size}. The \textit{left} panel assumes the \citet{chen17} empirical mass-radius relationship for the simulated planets and the currently measured masses for the known planets. The \textit{right} panel assumes that all the planets have Earth-like composition with their masses predicted by the formula given by \citet{zeng16} with a core mass fraction of 0.3.}
\end{center}
\end{figure*}

\section{Results \label{sec:results}}
\subsection{Statistical Sample \label{stat_sample}}
In Figure~\ref{fig:metric_size} we plot the 10-hour \textit{JWST}/NIRISS S/N from \citet{lou18} vs.\ the TSM from Equation~\ref{transmission_metric} for each planet size bin in the statistical sample. As can be seen, the analytic metric tracks the S/N calculated by \citet{lou18} with little scatter for targets with $m_{J}\,>\,9$. Simulated planets with particularly bright host stars exhibit a different slope in the relationship due to differences in the observational duty cycle with \textit{JWST} that our metric doesn't capture. In the \citet{lou18} simulations, stars with $m_J \sim 8.1$ \citep[depending on the stellar type, for more details see the University of Montreal \textit{JWST} website\footnote{\url{http://jwst.astro.umontreal.ca/?page_id=51}} and][]{beichman14} require the use of the bright star readout mode, and stars with $m_J\,<\,9$ start to have substantially lower duty cycles due to the limited number of reads possible before a reset of the detector is needed.

We did not attempt to correct for the mismatch between the \citet{lou18} results and the TSM for bright stars because \added{the TSM is intended as a general metric for the ranking of transmission spectroscopy targets for infrared observations.  A correction factor of the square root of the duty cycle can be applied to account for duty-cycle reductions for e.g bright stars with \textit{JWST}/NIRISS.}  Furthermore, systems with bright host stars have benefits that balance against their non-ideal nature for \textit{JWST} observations. For example, bright stars make RV mass measurements easier. Also, dedicated missions like \textit{ARIEL} have much smaller apertures than \textit{JWST} and will therefore only suffer reduced duty cycles for extremely bright stars. Bright stars are also typically preferred for ground-based high resolution observations due to the higher background in these data, the possibility of using adaptive optics systems to reduce slit losses, and the capability of these facilities and instruments to observe very bright stars without duty cycle penalties. Furthermore, higher efficiency read modes for \textit{JWST} observations of bright stars are currently being investigated \citep{batalha18}.

In Table~\ref{table:metric} we give the cutoff values of the TSM for the top quintile of terrestrial planets and the top 100 planets (sub-divided into 25-planet groupings) for the three largest size bins in the statistical sample. Both the large and small sub-Neptune bins have a plethora of targets that would yield high S/N atmospheric characterization with \textit{JWST}. However, the dramatic fall off in the TSM between the second and third quartiles for the sub-Jovian bin belies their relative scarcity at the orbital periods \textit{TESS} is sensitive to.  For this reason, we suggest selecting the top 50 sub-Jovian planets as the high-priority atmospheric characterization targets within that bin.

The values in Table~\ref{table:metric} can be used to prioritize observations aimed at confirming and measuring the masses of potential atmospheric characterization targets. For example, planets found early in the \textit{TESS} mission with smaller metric values than those reported in boldface in Table~\ref{table:metric} for the relevant bins could be confidently set aside in favor of better planets that will be found as the mission progresses. On the other hand, planets with metric values near the top should be seen as high priorities for follow-up efforts with the expectation that few other planets that are as good atmospheric characterization targets will be found later.


\begin{deluxetable*}{l c c c c c c c c c}
\tablecaption{Top 10 habitable zone planets --- empirical planet masses\label{table:metric_hz}}
\tablehead{ Name\tablenotemark{a} & TSM\tablenotemark{b} & \colhead{$M_{p}$} & \colhead{$R_{p}$} & \colhead{$M_{*}$} \vspace{-0.1cm} & \colhead{$R_{*}$} & \colhead{RV} &\colhead{$S$} & \colhead{$m_{V}$} & \colhead{$m_{J}$} \\ 
 & & \colhead{($M_{\oplus}$)} \vspace{+0.1cm} & \colhead{($R_{\oplus}$)} \vspace{+0.1cm} &  \colhead{($M_{\odot}$)} & \colhead{($R_{\odot}$)} & \colhead{(m/s)} & \colhead{($S_{\oplus}$)} & & }
\startdata
TESS-Sim  204  & 27.9  &  0.39  & 0.77  & 0.24  & 0.24  &  0.27  & 1.27  & 11.50  &  7.91 \\
TESS-Sim 1296  & 26.8  &  4.35  & 1.92  & 0.17  & 0.17  &  4.04  & 1.00  & 13.90  & 10.00 \\
TESS-Sim 1804  & 26.5  &  4.42  & 1.94  & 0.16  & 0.16  &  3.64  & 0.45  & 14.30  &  9.78 \\
TESS-Sim 1308  & 23.2  &  2.83  & 1.49  & 0.25  & 0.25  &  1.80  & 1.15  & 11.60  &  7.97 \\
TESS-Sim  922  & 21.6  &  2.53  & 1.39  & 0.12  & 0.12  &  4.16  & 1.17  & 16.20  & 11.26 \\
TESS-Sim  405  & 19.4  &  3.11  & 1.58  & 0.38  & 0.38  &  1.33  & 1.61  & 10.10  &  6.85 \\
TESS-Sim  105  & 17.9  &  3.48  & 1.68  & 0.14  & 0.14  &  4.15  & 1.12  & 15.30  & 11.27 \\
TESS-Sim   48  & 17.3  &  4.64  & 1.99  & 0.16  & 0.16  &  4.38  & 0.64  & 15.00  & 11.10 \\
TESS-Sim 1244  & 16.8  &  3.99  & 1.82  & 0.16  & 0.16  &  4.45  & 1.79  & 15.30  & 11.34 \\
TESS-Sim  991  & 15.8  &  3.67  & 1.74  & 0.16  & 0.16  &  2.97  & 0.39  & 14.40  & 10.53 \\
 \midrule
LHS\,1140b & 9.8  &  6.98 & 1.73 & 0.15 & 0.19 & 5.42 & 0.39 & 14.15 & 9.61  \\
TRAPPIST-1f & 23.3 & 0.69 & 1.05 & 0.09 & 0.12 & 1.05 & 0.36 & 18.80 & 11.35 \\
\enddata
\tablenotetext{a}{Planet names from the simulated \citet{sul15} \textit{TESS} catalog.}
\tablenotetext{b}{Scale factor = 0.167, calculated for the small temperate sample.}
\end{deluxetable*}

\begin{deluxetable*}{l c c c c c c c c c}
\tablecaption{Top 10 habitable zone planets --- rocky planet masses \label{table:metric_hz_rocky}}
\tablehead{ Name\tablenotemark{a} & TSM\tablenotemark{b} & \colhead{$M_{p}$} & \colhead{$R_{p}$} & \colhead{$M_{*}$} \vspace{-0.1cm} & \colhead{$R_{*}$} & \colhead{RV} &\colhead{$S$} & \colhead{$m_{V}$} & \colhead{$m_{J}$} \\ 
 & & \colhead{($M_{\oplus}$)} \vspace{+0.1cm} & \colhead{($R_{\oplus}$)} \vspace{+0.1cm} &  \colhead{($M_{\odot}$)} & \colhead{($R_{\odot}$)} & \colhead{(m/s)} & \colhead{($S_{\oplus}$)} & & }
\startdata 
TESS-Sim  204  & 28.2  &  0.38  & 0.77  & 0.24  & 0.24  &  0.27  & 1.27  & 11.50  &  7.91 \\
TESS-Sim  922  & 16.2  &  3.36  & 1.39  & 0.12  & 0.12  &  5.52  & 1.17  & 16.20  & 11.26 \\
TESS-Sim 1308  & 15.1  &  4.34  & 1.49  & 0.25  & 0.25  &  2.76  & 1.15  & 11.60  &  7.97 \\
TESS-Sim  405  & 11.2  &  5.40  & 1.58  & 0.38  & 0.38  &  2.31  & 1.61  & 10.10  &  6.85 \\
TESS-Sim 1878  & 11.0  &  1.32  & 1.08  & 0.14  & 0.14  &  1.44  & 0.78  & 15.40  & 11.37 \\
TESS-Sim 1296  & 10.5  & 11.10  & 1.92  & 0.17  & 0.17  & 10.30  & 1.00  & 13.90  & 10.00 \\
TESS-Sim 1804  & 10.2  & 11.53  & 1.94  & 0.16  & 0.16  &  9.49  & 0.45  & 14.30  &  9.78 \\
TESS-Sim   45  &  9.9  &  1.19  & 1.05  & 0.16  & 0.16  &  1.01  & 0.62  & 14.80  & 10.91 \\
TESS-Sim 1292  &  9.6  &  4.68  & 1.52  & 0.25  & 0.25  &  3.14  & 1.35  & 12.70  &  9.06 \\
TESS-Sim  105  &  9.2  &  6.77  & 1.68  & 0.14  & 0.14  &  8.07  & 1.12  & 15.30  & 11.27 \\
\midrule
LHS\,1140b & 9.15  &  7.50 & 1.73 & 0.15 & 0.19 & 5.83 & 0.39 & 14.15 & 9.61  \\
TRAPPIST-1f & 13.7 &  1.05 & 1.05 & 0.09 & 0.12 & 1.56 & 0.36 & 18.80 & 11.35 \\
\enddata
\tablenotetext{a}{Planet names from the simulated \citet{sul15} \textit{TESS} catalog.}
\tablenotetext{b}{Scale factor = 0.167, calculated for the small temperate sample.}
\end{deluxetable*}

\subsection{Small Temperate Sample \label{sm_tem}}
The results for the planets in and near the habitable zone are shown in the left panel of Figure~\ref{fig:metric_hz}. The analytic TSM performs very well for these simulated planets compared to the \citet{lou18} results, again with the exception of the small number of planets orbiting stars brighter than $m_{J}\,=\,9$. The TSM values and planet parameters for the top 10 simulated planets are given in Table~\ref{table:metric_hz}, benchmarked against the known planets LHS 1140b \citep{dittmann17,ment18} and TRAPPIST-1f \citep{gillon17,grimm18}. \added{For the simulated \textit{TESS} planets we estimate the stellar masses assuming a one-to-one relationship between stellar radius and mass, as indicated by \citet{boyajian12}, and compute the radial velocity signal consistent with our planet masses.}

The \citet{chen17} masses assumed for this study imply a significant volatile component for the larger planets in the small temperate sample, and thus the observability of these planets in the context of habitability may be overestimated compared to that of the smaller planets. Therefore, we repeat our analysis for this sample with a re-scaling of the \citet{lou18} results and the TSM assuming all the planets have Earth-like composition. Here we estimate the planet masses from their radii using the formula given by \citet{zeng16} with a core mass fraction of 0.3 (i.e., $M_{p}$\,=\,0.993\,$R_{p}^{3.7}$, where $M_{p}$ and $R_{p}$ are in Earth units). We also recompute the metric for the planets LHS\,1140b and TRAPPIST-1f with the same assumption of Earth-like composition. The results of this calculation are shown in the right panel of Figure~\ref{fig:metric_hz} and the parameters for the recomputed top 10 simulated planets are given in Table~\ref{table:metric_hz_rocky}. 

As previously pointed out by \citet{lou18}, although \textit{TESS} will indeed find small planets in the habitable zones of their host stars, the current expectation is that only a few of these planets will be as good or better targets for atmospheric characterization than the currently known planets. However, as also pointed out by \citet{lou18}, this prediction hinges critically on the assumed frequency of small planets around very small stars (note that all the host stars in Tables~\ref{table:metric_hz} and \ref{table:metric_hz_rocky} are late M dwarfs). The detection of systems like LHS\,1140 and TRAPPIST-1 with ground-based instruments suggests that perhaps the assumed occurrence rates in this regime are underestimated and that \textit{TESS} will find more of these nearby systems. Also, further photometric monitoring to search for additional transiting planets in \textit{TESS}-discovered systems may boost the yield of potentially habitable planets \citep{ballard18}.

Based on the values reported in Tables \ref{table:metric_hz} and \ref{table:metric_hz_rocky} we therefore propose that a TSM value of $\sim$10 (assuming either the empirical mass-radius relation or rocky composition for all) represents a good starting threshold for evaluating the atmospheric observability of potentially habitable planets identified with \textit{TESS}. Small planet candidates receiving Earth-like insolation and having TSM values substantially larger than this cutoff should be high priority targets for follow-up efforts in the context of future atmospheric characterization. Planets discovered early in the mission that are near or below this threshold value should only become high priority targets for follow-up observations for reasons other than atmospheric observability (e.g., exploring the mass-radius relationship for temperate planets). Ultimately, the TSM cutoff values for prioritization for each of the samples should be re-evaluated once we get a handle on the actual \textit{TESS} yield, say after the first year of the mission.

\begin{deluxetable*}{l c c c c c c c c c c c c}[t]
\tablecaption{Top emission spectroscopy targets \label{table:esm}}
\tablehead{ Name\tablenotemark{a} & ESM & TSM\tablenotemark{b} & \colhead{$M_{p}$} & \colhead{$R_{p}$} & \colhead{$T_{p}$} \vspace{-0.1cm} & \colhead{$P_{orb}$} & \colhead{$M_{*}$} & \colhead{$R_{*}$} & \colhead{$T_{*}$} &\colhead{RV} & \colhead{$m_{V}$} & \colhead{$m_{K}$} \\ 
 & & & \colhead{($M_{\oplus}$)} \vspace{+0.1cm} & \colhead{($R_{\oplus}$)} \vspace{+0.1cm} &  \colhead{(K)} & \colhead{(days)} & \colhead{($M_{\odot}$)} & \colhead{($R_{\odot}$)} & \colhead{(K)} & \colhead{(m/s)} & & }
\startdata
TESS-Sim  284  & 112.5  & 107.4  &  2.63  & 1.43  &  980 & 0.6 & 0.23  & 0.23  & 2560 &  5.36  & 13.10  &  6.80 \\
TESS-Sim 1763  &  29.8  &  49.8  &  2.46  & 1.37  & 1749 & 1.1 & 0.81  & 0.81  & 5190 &  1.75  &  6.47  &  4.47 \\
TESS-Sim 1476  &  26.8  &  77.9  &  0.28  & 1.38  &  661 & 2.3 & 0.29  & 0.29  & 3450 &  2.70  &  9.70  &  5.54 \\
TESS-Sim   21  &  23.6  &  35.9  &  2.53  & 1.49  & 1672 & 1.0 & 0.70  & 0.70  & 5030 &  2.27  &  8.00  &  5.85 \\
TESS-Sim 1855  &  14.7  &  23.6  &  1.31  & 1.35  &  990 & 1.1 & 0.42  & 0.42  & 3640 &  2.61  & 11.30  &  7.36 \\
TESS-Sim 1957  &  11.7  &  25.2  &  2.44  & 1.14  & 1815 & 1.0 & 0.81  & 0.81  & 5160 &  1.14  &  7.84  &  5.82 \\
TESS-Sim 1745  &  11.5  &  23.6  &  2.01  & 1.09  &  994 & 0.5 & 0.24  & 0.24  & 3340 &  2.69  & 14.00  &  9.59 \\
TESS-Sim 1421  &  10.8  &  15.8  &  2.26  & 1.40  &  950 & 0.6 & 0.27  & 0.27  & 3360 &  4.55  & 14.50  & 10.10 \\
TESS-Sim  858  &  10.5  &  12.4  &  1.23  & 1.48  & 1210 & 0.6 & 0.41  & 0.41  & 3590 &  3.87  & 13.60  &  9.53 \\
TESS-Sim 1255  &   9.9  &  22.4  &  0.39  & 1.22  &  697 & 0.6 & 0.14  & 0.14  & 2870 &  5.70  & 16.40  & 11.30 \\
TESS-Sim  675  &   9.6  &  14.1  &  1.18  & 1.47  &  905 & 0.6 & 0.21  & 0.21  & 3270 &  6.12  & 16.20  & 11.60 \\
TESS-Sim 1926  &   9.0  &  13.1  &  1.93  & 1.38  & 1028 & 0.8 & 0.37  & 0.37  & 3520 &  3.31  & 13.50  &  9.34 \\
TESS-Sim 1340  &   8.4  &  12.0  &  2.39  & 1.44  & 1084 & 1.1 & 0.48  & 0.48  & 3760 &  2.71  & 12.40  &  8.60 \\
TESS-Sim  289  &   8.2  &   9.9  &  2.83  & 1.49  & 1221 & 0.7 & 0.47  & 0.47  & 3740 &  3.37  & 13.30  &  9.48 \\
TESS-Sim   90  &   8.2  &  13.4  &  1.42  & 1.34  & 1007 & 1.4 & 0.48  & 0.48  & 3770 &  2.23  & 11.80  &  8.00 \\
TESS-Sim  419  &   8.2  &  15.9  &  2.58  & 1.09  & 1111 & 0.6 & 0.38  & 0.38  & 3530 &  1.90  & 12.90  &  8.81 \\
TESS-Sim 1780  &   8.2  &  11.6  &  2.56  & 1.26  & 1072 & 0.7 & 0.36  & 0.36  & 3500 &  3.08  & 13.60  &  9.48 \\
TESS-Sim  884  &   7.6  &   9.1  &  2.36  & 1.32  & 1298 & 0.5 & 0.43  & 0.43  & 3640 &  3.24  & 13.70  &  9.78 \\
TESS-Sim 1160  &   7.6  &   9.8  &  1.10  & 1.26  & 1265 & 0.5 & 0.43  & 0.43  & 3630 &  2.95  & 13.50  &  9.52 \\
TESS-Sim 1962  &   7.5  &   8.8  &  2.83  & 1.32  & 1342 & 0.5 & 0.46  & 0.46  & 3690 &  3.09  & 13.60  &  9.66 \\
\enddata
\tablenotetext{a}{Planet names from the simulated \citet{sul15} \textit{TESS} catalog.}
\tablenotetext{b}{TSM calculated with a scale factor of 0.190.}
\end{deluxetable*}

\subsection{Emission Sample}

Table~\ref{table:esm} identifies the set of 20 targets that have emission spectroscopy metric values larger than that of GJ\,1132b (ESM = 7.5).  Our benchmark planet GJ\,1132b is currently the best of the known small planets for secondary eclipse measurements at mid-IR wavelengths according to \citet{morley17}.  Even so, this planet is not guaranteed to be a straightforward target for emission spectroscopy with \textit{JWST}.  With a predicted S/N in the MIRI LRS bandpass-integrated secondary eclipse of 7.5, it will take at least two eclipses and photon-limited performance of the instrument to build up a ``white light'' S/N of greater than 10.  This should be seen as the minimum requirement for secondary eclipse \textit{spectroscopy}, which further reduces the S/N by dividing the observation into smaller wavelength intervals.  Furthermore, the single eclipse S/N for GJ\,1132b of 7.5 means that the secondary eclipse will easily be recoverable from a single epoch of observations --- an important consideration for planets whose secondary eclipse timing is not well constrained due to uncertainties on orbital eccentricities or ephemerides.  We therefore suggest that GJ\,1132b's ESM of 7.5 should be selected as the cutoff value in identifying the top emission spectroscopy small planets for \textit{JWST}. 

As expected, the planets in Table~\ref{table:esm} are distinguished by high $T_{eq}$, small $R_{*}$, and/or bright host stars.  All 20 planets have higher equilibrium temperatures than GJ\,1132b, affirming that this is likely to remain one of the best (if not the best) planets for thermal emission measurements with $T_{eq} < 600$ K in perpetuity.  Of the 20 planets identified in Table~\ref{table:esm}, the top 6 would be very challenging targets for \textit{JWST} due to the brightnesses of their host stars. Furthermore, the ESM likely overestimates the S/N of real observations for these systems because we neglect to consider the impact of the reduced duty cycle for very bright stars, as we did for the TSM \citep[although note that the MIRI detector is more efficient than the NIRISS detector,][]{batalha17}. And there is the issue that these bright systems require higher photon-limited precision that might not be obtainable in the face of instrument systematics. On the other hand, MIRI becomes background limited for stars fainter than $m_{J}\,=\,10$ \citep{batalha18}. Therefore, the S/N of real observations of some of the fainter targets in  Table~\ref{table:esm} would be lower than what is predicted from the ESM.

The other targets in Table~\ref{table:esm} have emission spectroscopy S/N values that are marginally better than that of GJ\,1132b.  As with the small temperate sample, the emission sample primarily identifies planets from within a region of parameter space where planet occurrence rates have high uncertainties (i.e. very small host stars and small planets on ultra-short period orbits).  We therefore caution that it is reasonable to expect the actual emission sample from \textit{TESS} will vary in size from our prediction of 20 planets.  

\section{Discussion and Conclusion \label{sec:concl}}

In summary, we have suggested simple, analytic metrics for determining which transiting exoplanets are the best targets for atmospheric characterization, with a focus on \textit{JWST} capabilities.  Applying these metrics specifically to the expected \textit{TESS} yield, we have determined appropriate cutoff values to identify planets that should be advanced expeditiously for RV follow-up and subsequent atmospheric investigations.  For the purpose of selecting easy-to-remember round numbers for the threshold transmission spectroscopy metric, based on the values in Table~\ref{table:metric}, we recommend that planets with TSM $> 10$ for $R_{p} < 1.5$~$R_{\oplus}$ and TSM $> 90$ for $1.5 < R_{p} < 10$~$R_{\oplus}$ be selected as high-quality atmospheric characterization targets among the \textit{TESS} planetary candidates.  We also recommend a threshold of TSM $= 10$ for putative habitable zone planets.  For emission spectroscopy of terrestrial planets, we recommend a threshold of ESM $= 7.5$.  Applying these cuts should result in $\sim 300$ new ideal targets for transmission spectroscopy investigations from the \textit{TESS} mission.  

\begin{figure}
\begin{center}
\includegraphics[scale=0.39]{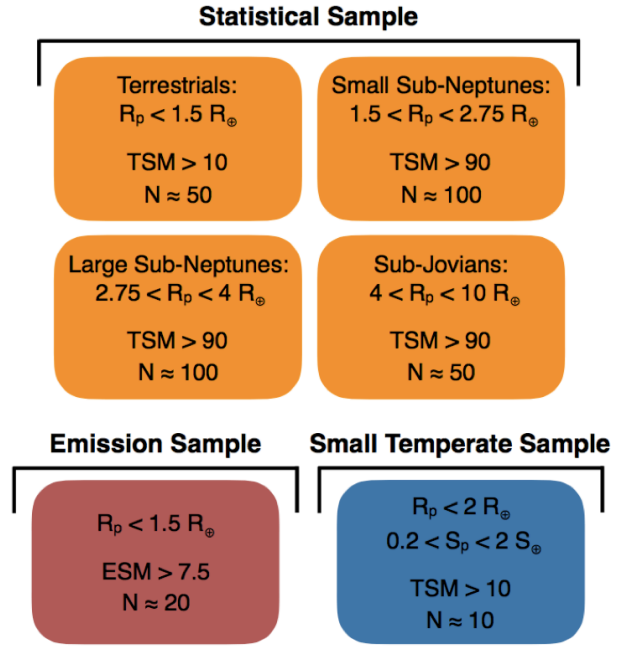}
\caption{\label{fig:summary} Summary of the properties and threshold metric values for the various atmospheric characterization samples described in this paper.}
\end{center}
\end{figure}

We review the various atmospheric characterization samples in Figure~\ref{fig:summary}, along with their metric selection criteria.  We note from Tables~\ref{table:metric_hz}, \ref{table:metric_hz_rocky}, and \ref{table:esm} that both the small temperate and emission samples are almost fully contained within the statistical sample.  Therefore, the selection criteria for the statistical sample should be seen as the primary mode for identifying appropriate \textit{JWST} atmospheric characterization targets from the \textit{TESS} returns. \added{Furthermore, the same TSM and ESM calculations can be used to identify high priority targets for exoplanet atmosphere studies with other facilities such as \textit{ARIEL} and ground-based ELTs, although in the latter case the threshold criteria may need to be revised to account for the stronger sensitivity of ground-based observations to the host star brightness.  The metric calculations can also be applied to existing exoplanet candidates such as those from the $K2$ mission to identify high-quality atmospheric characterization targets that will rival those that are expected to be discovered by \textit{TESS}.}

\added{We have also repeated our analysis using the simulated \textit{TESS} catalog of \citet{barclay18}, to assess the impact of using an independent realization of the mission outcome. We find that applying the same set of threshold TSM criteria from Figure~\ref{fig:summary} also results in a statistical sample of 250-300 planets, although the radius distribution of those objects is somewhat altered.  The \citet{barclay18} catalog returns more sub-Jovians and fewer sub-Neptunes (from both size bins from the latter category).  The larger number of sub-Jovians likely results from the inclusion of the \textit{TESS} full frame images in generating the \citet{barclay18} catalog, whereas the main \citet{sul15} catalog only accounted for the 2-minute cadence targets.  Somewhat surprisingly, despite the known overestimation of the number of terrestrial planets in \citet{sul15}, the \citet{barclay18} catalog actually produces several \textit{more} terrestrials above our suggested TSM threshold value of 10.  That is to say that while \citet{barclay18} predict fewer overall terrestrial planets, they produce higher quality targets with respect to transmission spectroscopy observations.}

\added{In addition to the selection of top atmospheric characterization targets using the TSM and ESM threshold values, additional factors may play into the decision to further prioritize or de-prioritize individual targets.} Refinement of the sample is also advised based on factors such as expected RV amplitude, stellar activity level, high false positive likelihood (e.g. near-grazing transits), \textit{JWST} observability (i.e. prioritizing targets that lie within the \textit{JWST} continuous viewing zone), \added{and the precision to which ephemerides and other system parameters are known.  Furthermore, while our aim has been to develop a truly statistical sample of exoplanet atmospheres, we acknowledge the biases that will ultimately remain in the selected targets.  For example, the terrestrial and small sub-Neptune samples are heavily dominated by planets orbiting M stars because of their relatively larger transit depths, whereas the sub-Jovian sample is weighted toward Sun-like hosts owing to the intrinsic scarcity of such planets around smaller M dwarfs.  Targets that buck these trends in host star type should therefore also be prioritized more highly.}

\begin{figure*}[tp]
\begin{center}
\includegraphics[scale=0.44]{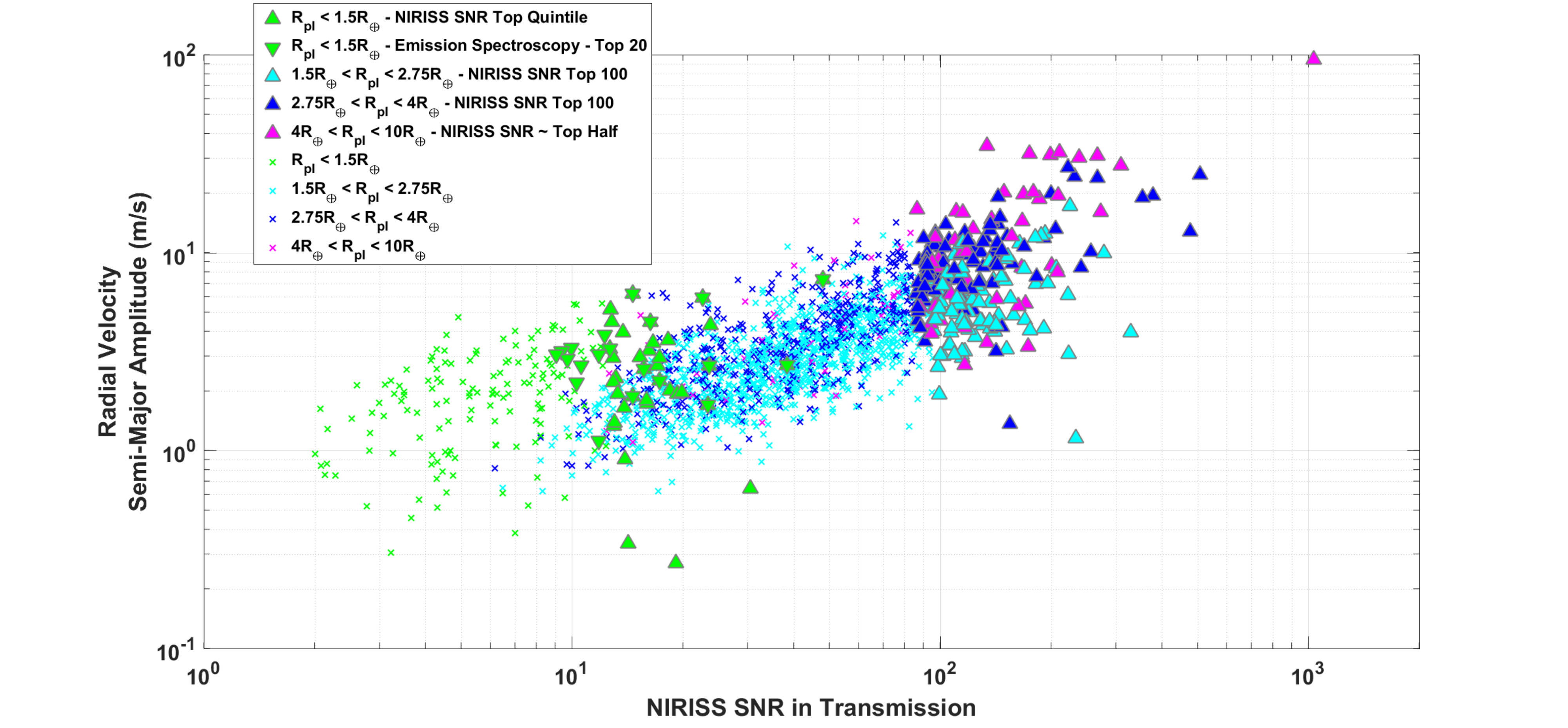}
\caption{\label{fig:RV_SN} RV semi-amplitude vs.~NIRISS S/N for the planets in the simulated \citet{sul15} \textit{TESS} catalog and the \citet{lou18} 10-hour S/N predictions.  Filled triangles denote the planets included in our transmission statistical sample using the threshold criteria from Table~\ref{table:metric}, upside down triangles indicate the planets included in our emission sample, and small x's denote targets that are disfavored for atmospheric characterization based on low expected S/N. (Note that the S/N values were calculated for a high-$\mu$ water-rich atmosphere for planets with $R_{p} < 1.5$~$R_{\oplus}$ and a low-$\mu$ hydrogen-rich atmosphere for $R_{p} > 1.5$~$R_{\oplus}$.)}
\end{center}
\end{figure*}

The similar metric cutoff values for the sub-Neptune and sub-Jovian planets reflects the fact that the \citet{lou18} simulation predicts similar S/N across a wide range of planetary sizes due to their adoption of the same atmospheric mean molecular weight and not accounting for the potential impact of clouds, which admittedly are difficult to impossible to accurately predict \textit{a priori}. \citet{crossfield17} have pointed out that smaller and cooler exoplanets empirically have smaller features in their transmission spectra relative to expectations. This suggests that such planets have higher mean molecular weight atmospheres and/or an increased prevalence of high altitude aerosols. Both of these phenomena would reduce the expected S/N from the nominal calculations of \citet{lou18}. So while the TSM is useful for prioritizing targets, the allocation of telescope time for atmospheric characterization will need to be carefully matched to the specific planets to be observed and the scientific objectives of the program.

The predicted RV semi-amplitudes of the simulated \textit{TESS} planets in the statistical sample are shown in Figure~\ref{fig:RV_SN} \added{(note that the stellar masses and resulting radial velocity signals are calculated as described in \S \ref{sm_tem})}. The bulk of the RV signals for the best targets identified in this study range between 1 -- 10\,m\,s$^{-1}$, which is within reach for many instruments that already exist or are currently under construction. However, it is worth acknowledging that currently only $\sim$250 known planets have both masses and radii measured to 10\% or better precision. Therefore, the goal laid out here of RV follow up for hundreds of new \textit{TESS} planets, most of which have $R_{p} < 4$~$R_{\oplus}$, may seem overly ambitious given the resource-intensive nature of RV observations. In addition, each of the small temperate planets listed in Tables~\ref{table:metric_hz} and \ref{table:metric_hz_rocky} will be very challenging for RV mass measurements owing to the small signals and relative faintness of the host stars.

Despite the obvious challenges, we propose that a large-scale effort to confirm and precisely measure the masses of hundreds of planets detected by \textit{TESS} is well justified. A key strength of exoplanet studies is the chance to perform statistical investigations that are not possible with the limited sample of planets in the Solar System. We are of the opinion that the study of exoplanet atmospheres is no different in this regard than studies of planetary frequency \citep{bean17b}. Furthermore, the study of a large sample of sub-Neptune exoplanet atmospheres is especially important because these planets do not exist in our Solar System, and therefore no well-studied benchmark objects exist. We also argue that a larger sample of sub-Neptune planets is needed than for giant planets due to the higher degree of diversity expected for these atmospheres in terms of their bulk compositions, which is a natural outcome of our proposed \textit{TESS} follow-up strategy.

Preliminary results suggest that measuring masses for $\sim$300 of the best \textit{TESS} planets for atmospheric characterization would require approximately 400 nights of observing time \citep{cloutier18}. Ultimately, we are optimistic that the large number of high-precision RV instruments expected to come online within the next few years \citep{fischer16,wright17} will bring the goal of dramatically expanding the sample of atmospheric characterization targets within reach.

\acknowledgments

The work of E.M.-R.K. was supported by the National Science Foundation under Grant No.\ 1654295 and by the Research Corporation for Science Advancement through their Cottrell Scholar program.  J.L.B.\ acknowledges support from the David and Lucile Packard Foundation and NASA through STScI grants GO-14792 and 14793. D.R.L.\ acknowledges support from NASA Headquarters under the NASA Earth and Space Science Fellowship (NESSF) Program - Grant NNX16AP52H. D.\ Deming acknowledges support from the TESS mission.  Part of the research was carried out at the Jet Propulsion Laboratory, California Institute of Technology, under contract with the National Aeronautics and Space Administration.  D.D.B.K.\ was supported by a James McDonnell Foundation postdoctoral fellowship. J.K.B. is supported by a Royal Astronomical Society Research Fellowship.  T.B. and E.V.Q. are grateful for support from GSFC Sellers Exoplanet Environments Collaboration (SEEC).  D.C. acknowledges support from the John Templeton Foundation. The opinions expressed here are those of the authors and do not necessarily reflect the views of John Templeton Foundation.  D.\ Dragomir acknowledges support provided by NASA through Hubble Fellowship grant HST-HF2-51372.001-A awarded by the Space Telescope Science Institute, which is operated by the Association of Universities for Research in Astronomy, Inc., for NASA, under contract NAS5-26555.  N.N. was partly supported by JSPS KAKENHI Grant Number JP18H01265 and JST PRESTO Grant Number JPMJPR1775. E.L.S. acknowledges funding from NASA Habitable Worlds grant  NNX16AB62G.  C.v.E.\ acknowledges funding for the Stellar Astrophysics Centre, provided by The Danish National Research Foundation (Grant agreement no.: DNRF106).

\bibliography{ms}

\begin{thebibliography}{}
\expandafter\ifx\csname natexlab\endcsname\relax\def\natexlab#1{#1}\fi
\providecommand{\url}[1]{\href{#1}{#1}}

\bibitem[{{Ballard}(2018)}]{ballard18}
{Ballard}, S. 2018, ArXiv e-prints, arXiv:1801.04949

\bibitem[{{Barclay} {et~al.}(2018){Barclay}, {Pepper}, \&
  {Quintana}}]{barclay18}
{Barclay}, T., {Pepper}, J., \& {Quintana}, E.~V. 2018, ArXiv e-prints,
  arXiv:1804.05050

\bibitem[{{Batalha} {et~al.}(2017{\natexlab{a}}){Batalha}, {Kempton}, \&
  {Mbarek}}]{bat17}
{Batalha}, N.~E., {Kempton}, E.~M.-R., \& {Mbarek}, R. 2017{\natexlab{a}},
  \apjl, 836, L5

\bibitem[{{Batalha} {et~al.}(2018){Batalha}, {Lewis}, {Line}, {Valenti}, \&
  {Stevenson}}]{batalha18}
{Batalha}, N.~E., {Lewis}, N.~K., {Line}, M.~R., {Valenti}, J., \& {Stevenson},
  K. 2018, \apjl, 856, L34

\bibitem[{{Batalha} \& {Line}(2017)}]{batalha17c}
{Batalha}, N.~E., \& {Line}, M.~R. 2017, \aj, 153, 151

\bibitem[{{Batalha} {et~al.}(2017{\natexlab{b}}){Batalha}, {Mandell},
  {Pontoppidan}, {Stevenson}, {Lewis}, {Kalirai}, {Earl}, {Greene}, {Albert},
  \& {Nielsen}}]{batalha17}
{Batalha}, N.~E., {Mandell}, A., {Pontoppidan}, K., {et~al.}
  2017{\natexlab{b}}, \pasp, 129, 064501

\bibitem[{{Bean} \& {FINESSE Science Team}(2017)}]{bean17}
{Bean}, J., \& {FINESSE Science Team}. 2017, in American Astronomical Society
  Meeting Abstracts, Vol. 229, American Astronomical Society Meeting Abstracts,
  301.08

\bibitem[{{Bean} {et~al.}(2017){Bean}, {Abbot}, \& {Kempton}}]{bean17b}
{Bean}, J.~L., {Abbot}, D.~S., \& {Kempton}, E.~M.-R. 2017, \apjl, 841, L24

\bibitem[{{Beichman} {et~al.}(2014){Beichman}, {Benneke}, {Knutson}, {Smith},
  {Lagage}, {Dressing}, {Latham}, {Lunine}, {Birkmann}, {Ferruit}, {Giardino},
  {Kempton}, {Carey}, {Krick}, {Deroo}, {Mandell}, {Ressler}, {Shporer},
  {Swain}, {Vasisht}, {Ricker}, {Bouwman}, {Crossfield}, {Greene}, {Howell},
  {Christiansen}, {Ciardi}, {Clampin}, {Greenhouse}, {Sozzetti}, {Goudfrooij},
  {Hines}, {Keyes}, {Lee}, {McCullough}, {Robberto}, {Stansberry}, {Valenti},
  {Rieke}, {Rieke}, {Fortney}, {Bean}, {Kreidberg}, {Ehrenreich}, {Deming},
  {Albert}, {Doyon}, \& {Sing}}]{beichman14}
{Beichman}, C., {Benneke}, B., {Knutson}, H., {et~al.} 2014, \pasp, 126, 1134

\bibitem[{{Berta-Thompson} {et~al.}(2015){Berta-Thompson}, {Irwin},
  {Charbonneau}, {Newton}, {Dittmann}, {Astudillo-Defru}, {Bonfils}, {Gillon},
  {Jehin}, {Stark}, {Stalder}, {Bouchy}, {Delfosse}, {Forveille}, {Lovis},
  {Mayor}, {Neves}, {Pepe}, {Santos}, {Udry}, \& {W{\"u}nsche}}]{berta15}
{Berta-Thompson}, Z.~K., {Irwin}, J., {Charbonneau}, D., {et~al.} 2015, \nat,
  527, 204

\bibitem[{{Bouma} {et~al.}(2017){Bouma}, {Winn}, {Kosiarek}, \&
  {McCullough}}]{bouma17}
{Bouma}, L.~G., {Winn}, J.~N., {Kosiarek}, J., \& {McCullough}, P.~R. 2017,
  ArXiv e-prints, arXiv:1705.08891

\bibitem[{{Boyajian} {et~al.}(2012){Boyajian}, {von Braun}, {van Belle},
  {McAlister}, {ten Brummelaar}, {Kane}, {Muirhead}, {Jones}, {White},
  {Schaefer}, {Ciardi}, {Henry}, {L{\'o}pez-Morales}, {Ridgway}, {Gies}, {Jao},
  {Rojas-Ayala}, {Parks}, {Sturmann}, {Sturmann}, {Turner}, {Farrington},
  {Goldfinger}, \& {Berger}}]{boyajian12}
{Boyajian}, T.~S., {von Braun}, K., {van Belle}, G., {et~al.} 2012, \apj, 757,
  112

\bibitem[{{Chen} \& {Kipping}(2017)}]{chen17}
{Chen}, J., \& {Kipping}, D. 2017, \apj, 834, 17

\bibitem[{{Cloutier} {et~al.}(2018){Cloutier}, {Doyon}, {Bouchy}, \&
  {H{\'e}brard}}]{cloutier18}
{Cloutier}, R., {Doyon}, R., {Bouchy}, F., \& {H{\'e}brard}, G. 2018, \aj, 156,
  82

\bibitem[{{Cowan} {et~al.}(2015){Cowan}, {Greene}, {Angerhausen}, {Batalha},
  {Clampin}, {Col{\'o}n}, {Crossfield}, {Fortney}, {Gaudi}, {Harrington},
  {Iro}, {Lillie}, {Linsky}, {Lopez-Morales}, {Mandell}, \&
  {Stevenson}}]{cowan15}
{Cowan}, N.~B., {Greene}, T., {Angerhausen}, D., {et~al.} 2015, \pasp, 127, 311

\bibitem[{{Crossfield} \& {Kreidberg}(2017)}]{crossfield17}
{Crossfield}, I.~J.~M., \& {Kreidberg}, L. 2017, \aj, 154, 261

\bibitem[{{Dittmann} {et~al.}(2017){Dittmann}, {Irwin}, {Charbonneau},
  {Bonfils}, {Astudillo-Defru}, {Haywood}, {Berta-Thompson}, {Newton},
  {Rodriguez}, {Winters}, {Tan}, {Almenara}, {Bouchy}, {Delfosse}, {Forveille},
  {Lovis}, {Murgas}, {Pepe}, {Santos}, {Udry}, {W{\"u}nsche}, {Esquerdo},
  {Latham}, \& {Dressing}}]{dittmann17}
{Dittmann}, J.~A., {Irwin}, J.~M., {Charbonneau}, D., {et~al.} 2017, \nat, 544,
  333

\bibitem[{{Dressing} \& {Charbonneau}(2015)}]{dressing15}
{Dressing}, C.~D., \& {Charbonneau}, D. 2015, \apj, 807, 45

\bibitem[{{Fischer} {et~al.}(2016){Fischer}, {Anglada-Escude}, {Arriagada},
  {Baluev}, {Bean}, {Bouchy}, {Buchhave}, {Carroll}, {Chakraborty}, {Crepp},
  {Dawson}, {Diddams}, {Dumusque}, {Eastman}, {Endl}, {Figueira}, {Ford},
  {Foreman-Mackey}, {Fournier}, {F{\H u}r{\'e}sz}, {Gaudi}, {Gregory},
  {Grundahl}, {Hatzes}, {H{\'e}brard}, {Herrero}, {Hogg}, {Howard}, {Johnson},
  {Jorden}, {Jurgenson}, {Latham}, {Laughlin}, {Loredo}, {Lovis}, {Mahadevan},
  {McCracken}, {Pepe}, {Perez}, {Phillips}, {Plavchan}, {Prato}, {Quirrenbach},
  {Reiners}, {Robertson}, {Santos}, {Sawyer}, {Segransan}, {Sozzetti},
  {Steinmetz}, {Szentgyorgyi}, {Udry}, {Valenti}, {Wang}, {Wittenmyer}, \&
  {Wright}}]{fischer16}
{Fischer}, D.~A., {Anglada-Escude}, G., {Arriagada}, P., {et~al.} 2016, \pasp,
  128, 066001

\bibitem[{{Fortney} {et~al.}(2013){Fortney}, {Mordasini}, {Nettelmann},
  {Kempton}, {Greene}, \& {Zahnle}}]{fortney13}
{Fortney}, J.~J., {Mordasini}, C., {Nettelmann}, N., {et~al.} 2013, \apj, 775,
  80

\bibitem[{{Fressin} {et~al.}(2013){Fressin}, {Torres}, {Charbonneau}, {Bryson},
  {Christiansen}, {Dressing}, {Jenkins}, {Walkowicz}, \& {Batalha}}]{fressin13}
{Fressin}, F., {Torres}, G., {Charbonneau}, D., {et~al.} 2013, \apj, 766, 81

\bibitem[{{Fulton} {et~al.}(2017){Fulton}, {Petigura}, {Howard}, {Isaacson},
  {Marcy}, {Cargile}, {Hebb}, {Weiss}, {Johnson}, {Morton}, {Sinukoff},
  {Crossfield}, \& {Hirsch}}]{fulton17}
{Fulton}, B.~J., {Petigura}, E.~A., {Howard}, A.~W., {et~al.} 2017, \aj, 154,
  109

\bibitem[{{Gillon} {et~al.}(2017){Gillon}, {Triaud}, {Demory}, {Jehin}, {Agol},
  {Deck}, {Lederer}, {de Wit}, {Burdanov}, {Ingalls}, {Bolmont}, {Leconte},
  {Raymond}, {Selsis}, {Turbet}, {Barkaoui}, {Burgasser}, {Burleigh}, {Carey},
  {Chaushev}, {Copperwheat}, {Delrez}, {Fernandes}, {Holdsworth}, {Kotze}, {Van
  Grootel}, {Almleaky}, {Benkhaldoun}, {Magain}, \& {Queloz}}]{gillon17}
{Gillon}, M., {Triaud}, A.~H.~M.~J., {Demory}, B.-O., {et~al.} 2017, \nat, 542,
  456

\bibitem[{{Girardi} {et~al.}(2005){Girardi}, {Groenewegen}, {Hatziminaoglou},
  \& {da Costa}}]{girardi05}
{Girardi}, L., {Groenewegen}, M.~A.~T., {Hatziminaoglou}, E., \& {da Costa}, L.
  2005, \aap, 436, 895

\bibitem[{{Grimm} {et~al.}(2018){Grimm}, {Demory}, {Gillon}, {Dorn}, {Agol},
  {Burdanov}, {Delrez}, {Sestovic}, {Triaud}, {Turbet}, {Bolmont}, {Caldas},
  {de Wit}, {Jehin}, {Leconte}, {Raymond}, {Van Grootel}, {Burgasser}, {Carey},
  {Fabrycky}, {Heng}, {Hernandez}, {Ingalls}, {Lederer}, {Selsis}, \&
  {Queloz}}]{grimm18}
{Grimm}, S.~L., {Demory}, B.-O., {Gillon}, M., {et~al.} 2018, ArXiv e-prints,
  arXiv:1802.01377

\bibitem[{{Howe} {et~al.}(2017){Howe}, {Burrows}, \& {Deming}}]{howe17}
{Howe}, A.~R., {Burrows}, A., \& {Deming}, D. 2017, \apj, 835, 96

\bibitem[{{Huang} {et~al.}(2018){Huang}, {Shporer}, {Dragomir}, {Fausnaugh},
  {Levine}, {Morgan}, {Nguyen}, {Ricker}, {Wall}, {Woods}, \&
  {Vanderspek}}]{huang18}
{Huang}, C.~X., {Shporer}, A., {Dragomir}, D., {et~al.} 2018, ArXiv e-prints,
  arXiv:1807.11129

\bibitem[{{Kane} {et~al.}(2009){Kane}, {Mahadevan}, {von Braun}, {Laughlin}, \&
  {Ciardi}}]{kane09}
{Kane}, S.~R., {Mahadevan}, S., {von Braun}, K., {Laughlin}, G., \& {Ciardi},
  D.~R. 2009, \pasp, 121, 1386

\bibitem[{{Kendrew} {et~al.}(2015){Kendrew}, {Scheithauer}, {Bouchet},
  {Amiaux}, {Azzollini}, {Bouwman}, {Chen}, {Dubreuil}, {Fischer}, {Glasse},
  {Greene}, {Lagage}, {Lahuis}, {Ronayette}, {Wright}, \& {Wright}}]{kendrew15}
{Kendrew}, S., {Scheithauer}, S., {Bouchet}, P., {et~al.} 2015, \pasp, 127, 623

\bibitem[{Koll \& Abbot(2015)}]{koll2015}
Koll, D. D.~B., \& Abbot, D.~S. 2015, The Astrophysical Journal, 802, 21

\bibitem[{Koll \& Abbot(2016)}]{koll2016}
---. 2016, The Astrophysical Journal, 825, 99

\bibitem[{{Kopparapu} {et~al.}(2013){Kopparapu}, {Ramirez}, {Kasting}, {Eymet},
  {Robinson}, {Mahadevan}, {Terrien}, {Domagal-Goldman}, {Meadows}, \&
  {Deshpande}}]{kopparapu13}
{Kopparapu}, R.~K., {Ramirez}, R., {Kasting}, J.~F., {et~al.} 2013, \apj, 765,
  131

\bibitem[{{Louie} {et~al.}(2018){Louie}, {Deming}, {Albert}, {Bouma}, {Bean},
  \& {Lopez-Morales}}]{lou18}
{Louie}, D.~R., {Deming}, D., {Albert}, L., {et~al.} 2018, \pasp, 130, 044401

\bibitem[{{Ment} {et~al.}(2018){Ment}, {Dittmann}, {Astudillo-Defru},
  {Charbonneau}, {Irwin}, {Bonfils}, {Murgas}, {Almenara}, {Forveille}, {Agol},
  {Ballard}, {Berta-Thompson}, {Bouchy}, {Cloutier}, {Delfosse}, {Doyon},
  {Dressing}, {Esquerdo}, {Haywood}, {Kipping}, {Latham}, {Lovis}, {Newton},
  {Pepe}, {Rodriguez}, {Santos}, {Tan}, {Udry}, {Winters}, \&
  {W{\"u}nsche}}]{ment18}
{Ment}, K., {Dittmann}, J.~A., {Astudillo-Defru}, N., {et~al.} 2018, ArXiv
  e-prints, arXiv:1808.00485

\bibitem[{{Miller-Ricci} {et~al.}(2009){Miller-Ricci}, {Seager}, \&
  {Sasselov}}]{mil09}
{Miller-Ricci}, E., {Seager}, S., \& {Sasselov}, D. 2009, \apj, 690, 1056

\bibitem[{{Mordasini} {et~al.}(2016){Mordasini}, {van Boekel}, {Molli{\`e}re},
  {Henning}, \& {Benneke}}]{mordasini16}
{Mordasini}, C., {van Boekel}, R., {Molli{\`e}re}, P., {Henning}, T., \&
  {Benneke}, B. 2016, \apj, 832, 41

\bibitem[{{Morgan} {et~al.}(2018){Morgan}, {Kerins}, {Awiphan}, {McDonald},
  {Hayes}, {Komonjinda}, {Mkritchian}, \& {Sanguansak}}]{morgan18}
{Morgan}, J., {Kerins}, E., {Awiphan}, S., {et~al.} 2018, ArXiv e-prints,
  arXiv:1802.05645

\bibitem[{{Morley} {et~al.}(2017){Morley}, {Kreidberg}, {Rustamkulov},
  {Robinson}, \& {Fortney}}]{morley17}
{Morley}, C.~V., {Kreidberg}, L., {Rustamkulov}, Z., {Robinson}, T., \&
  {Fortney}, J.~J. 2017, \apj, 850, 121

\bibitem[{{Ricker} {et~al.}(2015){Ricker}, {Winn}, {Vanderspek}, {Latham},
  {Bakos}, {Bean}, {Berta-Thompson}, {Brown}, {Buchhave}, {Butler}, {Butler},
  {Chaplin}, {Charbonneau}, {Christensen-Dalsgaard}, {Clampin}, {Deming},
  {Doty}, {De Lee}, {Dressing}, {Dunham}, {Endl}, {Fressin}, {Ge}, {Henning},
  {Holman}, {Howard}, {Ida}, {Jenkins}, {Jernigan}, {Johnson}, {Kaltenegger},
  {Kawai}, {Kjeldsen}, {Laughlin}, {Levine}, {Lin}, {Lissauer}, {MacQueen},
  {Marcy}, {McCullough}, {Morton}, {Narita}, {Paegert}, {Palle}, {Pepe},
  {Pepper}, {Quirrenbach}, {Rinehart}, {Sasselov}, {Sato}, {Seager},
  {Sozzetti}, {Stassun}, {Sullivan}, {Szentgyorgyi}, {Torres}, {Udry}, \&
  {Villasenor}}]{ricker15}
{Ricker}, G.~R., {Winn}, J.~N., {Vanderspek}, R., {et~al.} 2015, Journal of
  Astronomical Telescopes, Instruments, and Systems, 1, 014003

\bibitem[{{Rieke} {et~al.}(2015){Rieke}, {Wright}, {B{\"o}ker}, {Bouwman},
  {Colina}, {Glasse}, {Gordon}, {Greene}, {G{\"u}del}, {Henning}, {Justtanont},
  {Lagage}, {Meixner}, {N{\o}rgaard-Nielsen}, {Ray}, {Ressler}, {van Dishoeck},
  \& {Waelkens}}]{rieke15}
{Rieke}, G.~H., {Wright}, G.~S., {B{\"o}ker}, T., {et~al.} 2015, \pasp, 127,
  584

\bibitem[{Robinson \& Catling(2014)}]{robinson2014}
Robinson, T.~D., \& Catling, D.~C. 2014, Nature Geoscience, 7, 12

\bibitem[{{Rodler} \& {L{\'o}pez-Morales}(2014)}]{rodler14}
{Rodler}, F., \& {L{\'o}pez-Morales}, M. 2014, \apj, 781, 54

\bibitem[{{Schaefer} {et~al.}(2016){Schaefer}, {Wordsworth}, {Berta-Thompson},
  \& {Sasselov}}]{schaefer16}
{Schaefer}, L., {Wordsworth}, R.~D., {Berta-Thompson}, Z., \& {Sasselov}, D.
  2016, \apj, 829, 63

\bibitem[{{Snellen} {et~al.}(2013){Snellen}, {de Kok}, {le Poole}, {Brogi}, \&
  {Birkby}}]{snellen13}
{Snellen}, I.~A.~G., {de Kok}, R.~J., {le Poole}, R., {Brogi}, M., \& {Birkby},
  J. 2013, \apj, 764, 182

\bibitem[{{Stassun} {et~al.}(2018){Stassun}, {Corsaro}, {Pepper}, \&
  {Gaudi}}]{stassun18}
{Stassun}, K.~G., {Corsaro}, E., {Pepper}, J.~A., \& {Gaudi}, B.~S. 2018, \aj,
  155, 22

\bibitem[{{Sullivan} {et~al.}(2015){Sullivan}, {Winn}, {Berta-Thompson},
  {Charbonneau}, {Deming}, {Dressing}, {Latham}, {Levine}, {McCullough},
  {Morton}, {Ricker}, {Vanderspek}, \& {Woods}}]{sul15}
{Sullivan}, P.~W., {Winn}, J.~N., {Berta-Thompson}, Z.~K., {et~al.} 2015, \apj,
  809, 77

\bibitem[{{Thompson} {et~al.}(2018){Thompson}, {Coughlin}, {Hoffman},
  {Mullally}, {Christiansen}, {Burke}, {Bryson}, {Batalha}, {Haas},
  {Catanzarite}, {Rowe}, {Barentsen}, {Caldwell}, {Clarke}, {Jenkins}, {Li},
  {Latham}, {Lissauer}, {Mathur}, {Morris}, {Seader}, {Smith}, {Klaus},
  {Twicken}, {Van Cleve}, {Wohler}, {Akeson}, {Ciardi}, {Cochran}, {Henze},
  {Howell}, {Huber}, {Pr{\v s}a}, {Ram{\'{\i}}rez}, {Morton}, {Barclay},
  {Campbell}, {Chaplin}, {Charbonneau}, {Christensen-Dalsgaard}, {Dotson},
  {Doyle}, {Dunham}, {Dupree}, {Ford}, {Geary}, {Girouard}, {Isaacson},
  {Kjeldsen}, {Quintana}, {Ragozzine}, {Shabram}, {Shporer}, {Silva Aguirre},
  {Steffen}, {Still}, {Tenenbaum}, {Welsh}, {Wolfgang}, {Zamudio}, {Koch}, \&
  {Borucki}}]{thompson18}
{Thompson}, S.~E., {Coughlin}, J.~L., {Hoffman}, K., {et~al.} 2018, \apjs, 235,
  38

\bibitem[{{Tinetti} {et~al.}(2016){Tinetti}, {Drossart}, {Eccleston},
  {Hartogh}, {Heske}, {Leconte}, {Micela}, {Ollivier}, {Pilbratt}, {Puig},
  {Turrini}, {Vandenbussche}, {Wolkenberg}, {Pascale}, {Beaulieu}, {G{\"u}del},
  {Min}, {Rataj}, {Ray}, {Ribas}, {Barstow}, {Bowles}, {Coustenis}, {Coud{\'e}
  du Foresto}, {Decin}, {Encrenaz}, {Forget}, {Friswell}, {Griffin}, {Lagage},
  {Malaguti}, {Moneti}, {Morales}, {Pace}, {Rocchetto}, {Sarkar}, {Selsis},
  {Taylor}, {Tennyson}, {Venot}, {Waldmann}, {Wright}, {Zingales}, \&
  {Zapatero-Osorio}}]{tinetti16}
{Tinetti}, G., {Drossart}, P., {Eccleston}, P., {et~al.} 2016, in \procspie,
  Vol. 9904, Space Telescopes and Instrumentation 2016: Optical, Infrared, and
  Millimeter Wave, 99041X

\bibitem[{{Torres} {et~al.}(2011){Torres}, {Fressin}, {Batalha}, {Borucki},
  {Brown}, {Bryson}, {Buchhave}, {Charbonneau}, {Ciardi}, {Dunham}, {Fabrycky},
  {Ford}, {Gautier}, {Gilliland}, {Holman}, {Howell}, {Isaacson}, {Jenkins},
  {Koch}, {Latham}, {Lissauer}, {Marcy}, {Monet}, {Prsa}, {Quinn}, {Ragozzine},
  {Rowe}, {Sasselov}, {Steffen}, \& {Welsh}}]{torres11}
{Torres}, G., {Fressin}, F., {Batalha}, N.~M., {et~al.} 2011, \apj, 727, 24

\bibitem[{{Van Eylen} {et~al.}(2018){Van Eylen}, {Agentoft}, {Lundkvist},
  {Kjeldsen}, {Owen}, {Fulton}, {Petigura}, \& {Snellen}}]{vaneylen18}
{Van Eylen}, V., {Agentoft}, C., {Lundkvist}, M.~S., {et~al.} 2018, \mnras,
  479, 4786

\bibitem[{{Wright} \& {Robertson}(2017)}]{wright17}
{Wright}, J.~T., \& {Robertson}, P. 2017, Research Notes of the American
  Astronomical Society, 1, 51

\bibitem[{{Zellem} {et~al.}(2017){Zellem}, {Swain}, {Roudier}, {Shkolnik},
  {Creech-Eakman}, {Ciardi}, {Line}, {Iyer}, {Bryden}, {Llama}, \&
  {Fahy}}]{zellem17}
{Zellem}, R.~T., {Swain}, M.~R., {Roudier}, G., {et~al.} 2017, \apj, 844, 27

\bibitem[{{Zellem} {et~al.}(2018){Zellem}, {Fortney}, {Swain}, {Bryden},
  {Chapman}, {Cowan}, {Kataria}, {Kreidberg}, {Line}, {Moses}, {Parmentier},
  {Roudier}, \& {Stevenson}}]{zellem18}
{Zellem}, R.~T., {Fortney}, J.~J., {Swain}, M.~R., {et~al.} 2018, ArXiv
  e-prints, arXiv:1803.07163

\bibitem[{{Zeng} {et~al.}(2016){Zeng}, {Sasselov}, \& {Jacobsen}}]{zeng16}
{Zeng}, L., {Sasselov}, D.~D., \& {Jacobsen}, S.~B. 2016, \apj, 819, 127

\end{thebibliography}

\end{document}